\documentclass[11pt]{article} 
\usepackage[utf8]{inputenc}
\usepackage{multirow}
\usepackage{amsfonts}
\usepackage{amsmath}
\usepackage{amssymb}
\usepackage{amsthm}
\usepackage{caption}
\usepackage[dvipsnames]{xcolor}
    \definecolor{darkgreen}{rgb}{0,0.5,0}
    \definecolor{darkblue}{rgb}{0,0,0.6}
    \definecolor{purple}{rgb}{0.4,.2,0.7}
\usepackage[margin = 2.5cm]{geometry}
\usepackage{graphicx}
\usepackage[hyperfootnotes = false, colorlinks = true, linkcolor = darkblue, citecolor = purple]{hyperref}
\usepackage{subcaption}
\usepackage{ytableau}

\usepackage{comment}

\newcommand{\be}{\begin{equation}}
\newcommand{\ee}{\end{equation}}
\newcommand{\bea}{\begin{eqnarray}}
\newcommand{\eea}{\end{eqnarray}}
\def\la{\label}
\def\nref#1{(\ref{#1})}
\def\half{{1 \over 2 }}

\begin{document}

\thispagestyle{empty}
\begin{center}
    ~\vspace{5mm}

  \vskip 2cm 
  
   {\LARGE \bf 
       Soft Theorems in Matrix Theory
   }

   \vspace{0.5in}
     
   {\bf Aidan Herderschee and Juan Maldacena }

    \vspace{0.5in}

  Institute for Advanced Study,  Princeton, NJ 08540, USA

    \vspace{0.5in}

    \vspace{0.5in}
    

\end{center}

\vspace{0.5in}

\begin{abstract}
 
  \end{abstract}
 
  We show that the  Banks-Fischler-Shenker-Susskind matrix model for M-theory obeys the leading and subleading soft theorems expected from eleven-dimensional supergravity. 
The subleading soft theorem implies the amplitude is Lorentz symmetric. This is argued for general four point amplitudes, but only for restricted kinematics for five and higher point amplitudes.

\vspace{1in}

\pagebreak

\setcounter{tocdepth}{3}
{\hypersetup{linkcolor=black}\tableofcontents}

\section{Introduction  }  

The BFSS conjecture \cite{Banks:1996vh} states that the large $N$ limit of a certain matrix model gives the scattering amplitudes of massless particles in eleven dimensional M-theory. It is supposed to give a full non-perturbative description of flat space physics in eleven dimensions and is therefore an explicit example of flat space holography. The full Lorentz symmetry is not present in the matrix model for finite $N$, but is supposed to emerge in the large $N$ limit. In this paper, we derive the soft theorems, after we make some assumptions that we spell out below.  These soft theorems can then be used to argue for Lorentz invariance.

\begin{figure}[h]
    \begin{center}
     \includegraphics[scale=.3]{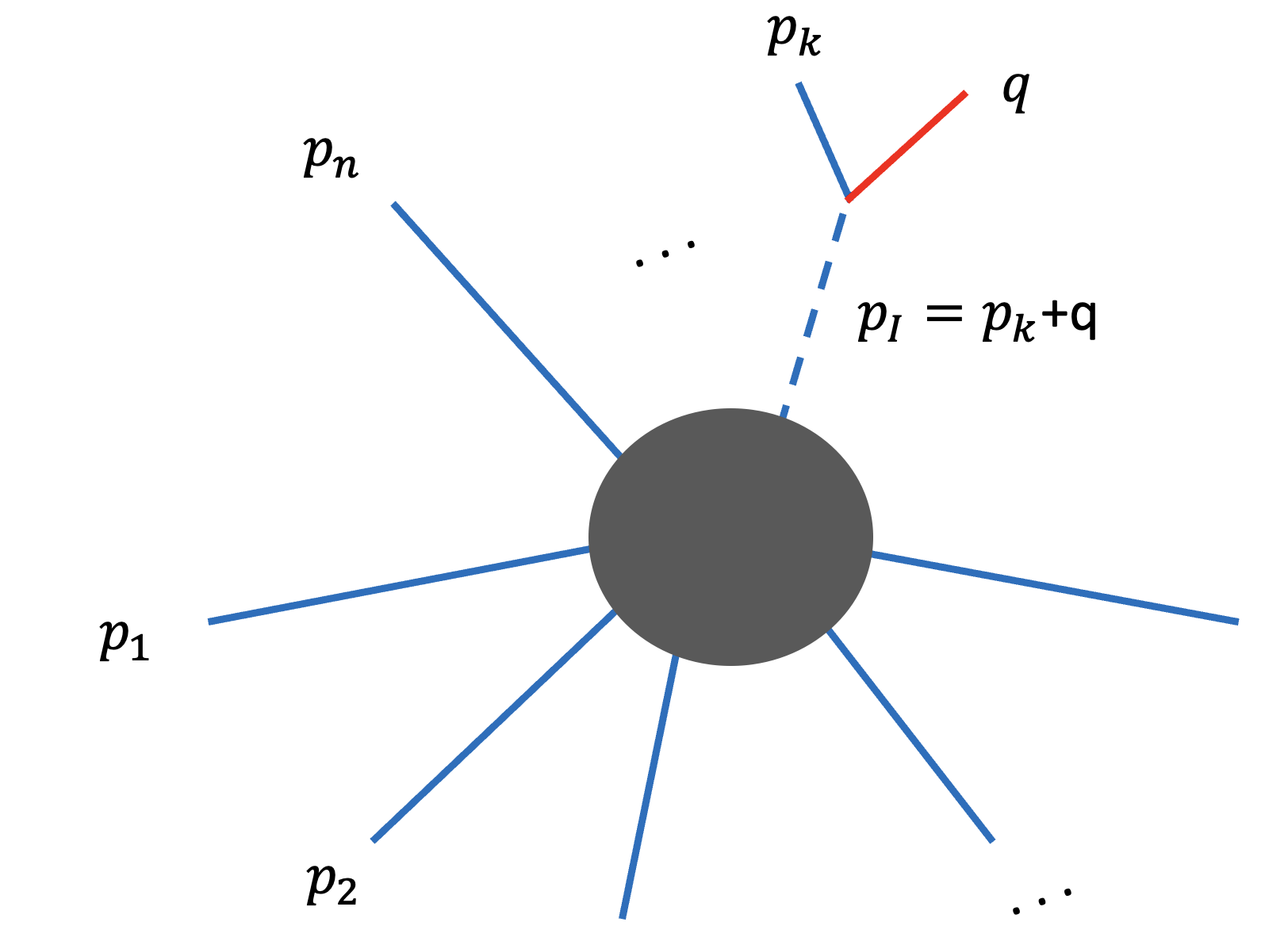} 
    \end{center}
     \caption{The soft limit depends on the amplitude having a factorization channel when the soft momentum and any of the external particles obeys $(p_k +q)^2 \to 0$. Crucially, the three point amplitude that appears here should be the usual one.   }
    \label{Soft}
 \end{figure}

In theories of gravity, we have soft theorems that indicate how amplitudes behave when one of the particles has a momentum $q_\mu$ that is taken to zero. 
Schematically, they have the form \cite{Weinberg:1965nx,Cachazo:2014fwa}
\be \la{SofTh}
{\cal A}_{n+1}(p_1, \cdots p_n , q)  \sim  ( S_{0} + S_1 ) {\cal A}_{n} (p_1, \cdots , p_n) + \cdots  ~,~~~~~{\rm as }~~ q\to 0 \ .
\ee 
The leading soft factor scales like $1/q$ and the second like $q^0$ but depends on the direction in which the limit is taken.   The dots indicate terms that are less singular when $q\to 0$.  These soft factors can be written as a sum of terms that involve the momenta of each of the $n$ particles and also involve the choice of an unphysical reference vector related to the description of the polarization of the soft graviton. Independence on this vector implies that the $n$ particle amplitude on the right hand side of  \nref{SofTh} should obey momentum conservation  and Lorentz and rotational invariance \cite{Weinberg:1965nx,Cachazo:2014fwa}. 
These soft theorems depend on the three point amplitude and on factorization properties of the $n+1$ point amplitudes, see figure \ref{Soft}. These soft theorems are now well understood at tree and loop level in gravitational theories, see e.g. \cite{Schwab:2014xua,Afkhami-Jeddi:2014fia,Kalousios:2014uva,Cachazo:2014dia,Bern:2014oka,Broedel:2014fsa,He:2014bga,Elvang:2016qvq,Sen:2017nim,Saha:2019tub,Yan:2023rjh}.

In this paper, we present an argument for the soft theorems in the BFSS matrix model. We were motivated in part by recent discussions of ``celestial holography'' where soft theorems play an important role \cite{Kapec:2014opa,Kapec:2015vwa,Strominger:2017zoo}. We wanted to understand whether the soft theorems hold in the explicit example of celestial holography, or flat space holography, that the BFSS model provides. Recent studies in this direction include \cite{Miller:2022fvc,Tropper:2023fjr}. Our discussion is an improvement because it gives a derivation of the soft limit theorems directly in the matrix model without referencing the eleven-dimensional theory, except for the assumption that a smooth large $N$ limit exists. The crucial input is the on-shell three-point amplitude with a soft momentum of non-zero $N_{\rm soft}$, which was derived from first principles in \cite{Herderschee:2023pza}. After taking the large $N$ limit, the soft momentum becomes a continuous variable $-q_- = { N_{\rm soft} \over R_-}$, making it possible to take derivatives with respect to $p_-^k$. These derivatives appear in a crucial term in the subleading soft factor, giving rise to the non-trivial Lorentz generators that are broken by the null compactification but should re-emerge in the large $N$ limit.

Though our main target is the BFSS model, we structure the paper in a way that highlights the origin of the soft theorems as following from a series of assumptions, regardless of the specific theory. We then discuss whether these assumptions are correct in the BFSS matrix model. 

In section \ref{GenSoft}, we discuss our assumptions for deriving the leading and subleading soft theorems. These assumptions are essentially the following. We assume that the external particles are massless and that their three point amplitudes are those of minimally coupled tree level gravity. We then assume that they factorize appropriately on poles when $(p_k + q )^2 \to 0$, see figure \ref{Soft},  and that possible branch cuts in the amplitude have a subleading dependence on $q$ in the $q\to 0$ limit. 
  In particular, we do not assume that the amplitudes are Lorentz invariant, except for the three-point amplitude. Using these assumptions we derive the soft theorem by performing a BCFW-like complex shift of the momenta. This shift introduces a vector that should not affect the final answer. This leads to the condition that the amplitudes are momentum conserving and Lorentz invariant \cite{Weinberg:1965nx,Cachazo:2014fwa}. 
  
  In section \ref{bfssmod}, we apply these ideas to the BFSS matrix model \cite{Banks:1996vh}. We make the non-trivial assumption that the large $N$ limit appearing in this conjecture gives a smooth function of $p^i_- \equiv -N_i/R_-$, in the limit $N_i , ~R_- \to \infty $ keeping $p_-^i$ fixed.  In other words, we simply assume that the limit of the matrix model scattering problem exists and that it is an analytic function (with poles, branch cuts, etc) of the momenta $p_-^{i} , \vec p^{\,\, i}$ of the $n$ particles. However, we do {\it not} assume that the amplitude is Lorentz invariant or that it coincides with gravity. 
 In \cite{Herderschee:2023pza}, we have shown from first principles that this limit indeed exists for the three-point amplitude, that it is Lorentz invariant, and that it coincides with the standard three-point function as implied by super-Poincare symmetry.  
 We then argue for the factorization properties assumed above. 
 We can then conclude that the amplitudes that we compute in the BFSS model are Lorentz invariant. 
 
Due to our technical limitations, we restricted the kinematics to lie in a four-dimensional subspace.\footnote{We use four-dimensional spinor helicity throughout this paper, which is reviewed in \cite{Elvang:2013cua}.} This is not a problem for the four-point amplitude, which can always be taken to lie in that subspace, but it does not describe the most general $n\geq 5$ amplitudes. We expect that the ideas in this paper could be easily extended to cover general kinematic configurations.     
  
 \section{General argument for Soft theorems } 
 \la{GenSoft}
 
Here we present an argument stating that any scattering amplitude of massless particles obeying a series of assumptions needs to also obey the soft theorems and be Lorentz invariant.  
 
 \subsection{Assumptions } 
 \la{Assumptions}

 We now list our assumptions for proving the leading and subleading soft theorems: 
  \begin{itemize}
 
 \item[1)] The amplitudes $A_n( p_\mu^i)$ are analytic functions of their arguments. This means that they are analytic for generic real momenta, but they can have poles or branch cuts as they are analytically continued in the momenta to complex momenta.  
 
 \item[2)] When one particle is soft, it has a pole as $(p_i + q)^2 \to 0$ which is of the form 
 \be \la{FacPo}
 {\cal A}_{n+1 } \sim {\cal A}_3 { 1 \over (p_i + q)^2 } {\cal A}_n ~,~~~~~{\rm for} ~~~(p_i + q)^2 \to 0
 \ee  
 where the amplitudes on the right-hand side are on-shell amplitudes evaluated at $p_I =p_i + q $, with
 $p_I^2 =0$. 
 
 \item[3)] Other singularities, such as branch cuts, behave as $q^{\alpha}$, for $\alpha>0$ when $q\to 0$. In our argument below, we will see more clearly how we use this condition and state more precisely what we need. 
 
 \item[4)] The three-point amplitude in \nref{FacPo} is the standard, relativistic three-point amplitude of a minimally-coupled graviton with another massless particle. In particular, note that the $n\geq 4 $ point amplitudes are not assumed to be Lorentz invariant. 
 	
 \end{itemize}
 These assumptions should hold for any minimally-coupled gravitational theory in $D>4$ dimensions. Assumption 3) is actually broken in four dimensional gravitational theories and the subleading soft theorem subsequently receives a loop correction \cite{Bern:2014oka,He:2014bga}.\footnote{It may seem strange that we are restricting to four dimensional kinematics while our assumptions only hold for gravitational theories in $D>4$ dimensions. The key point is that we are restricting only the external kinematics to lie in a four-dimensional sub-space. Momenta running in loops do not obey such a restriction.}

\subsection{The contour deformation argument using a BCFW shift } \la{BCFWsec}

We now present an argument that is in the spirit of   BCFW \cite{Britto:2004ap,Britto:2005fq}  as applied to the on-shell derivation of the soft factor in \cite{Arkani-Hamed:2008owk}. The argument is technically similar to that in \cite{Cachazo:2014fwa}, except that we do not take the deformed contour to infinity and argue that the derivation applies to the full amplitude, not just the tree-level part. For technical reasons, we restrict to four-dimensional kinematics, with momenta lying in a four-dimensional subspace of a higher-dimensional theory. Finally, we will consider the holomorphic soft limit \cite{Arkani-Hamed:2008owk,Cachazo:2014fwa}
\begin{equation}
|s\rangle \rightarrow \epsilon |s\rangle, \quad |s]\rightarrow |s]
\end{equation}
which is the same as the usual soft limit up to an overall rescaling of the amplitude.

 Since we have four-dimensional kinematics, we represent states in terms of spinor helicity variables. We consider an amplitude 
 \be 
 {\cal A}_{n+1}( s, 1, \cdots , n) 
 \ee 
 that depends on the spinor helicity variables $\lambda^i_{\alpha}$ and $\bar \lambda^i_{\alpha }$.   We will also strip off the momentum-conserving delta function. The first particle is the soft particle, denoted by $s$. We then perform a general momentum-conserving BCFW shift \cite{Britto:2005fq}  involving the soft particle and the particle $n$: 
 \be \la{BCFWShift}
 |s \rangle \to |s(z) \rangle = |s\rangle + z |n \rangle ~,~~~~~~~~~|n] \to |n(z)] = |n] - z |s] 
 \ee 
 where the  brackets $|i\rangle$ $| i]$ denote the spinor helicity variables  $\lambda^i_{\alpha}$ and $\bar \lambda^i_{\dot \beta }$ respectively. We do not touch the other variables. 

\begin{figure}[h]
    \begin{center}
     \includegraphics[scale=.3]{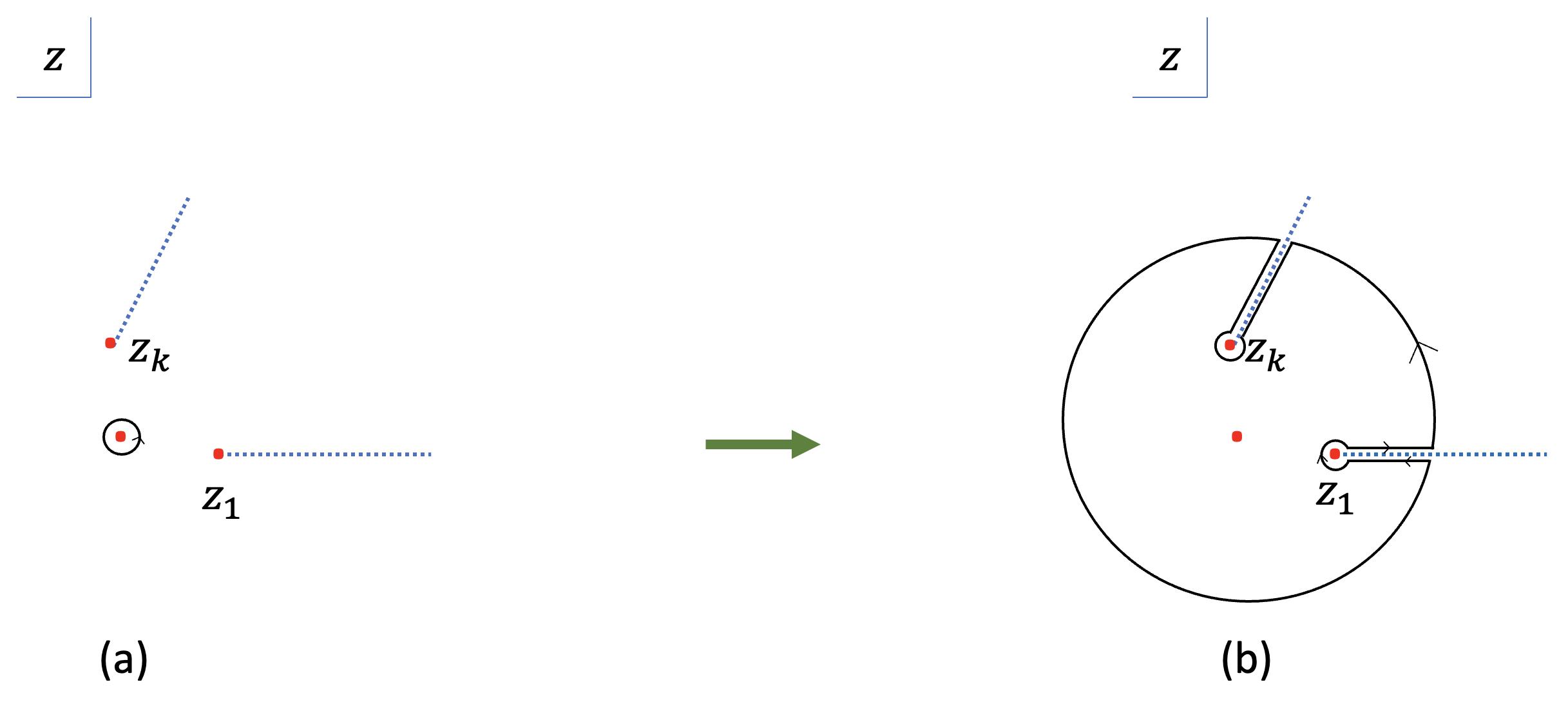} 
    \end{center}
     \caption{(a) The $z$ plane and the original contour around $z=0$. There are poles at $z_k$ and branch cuts that end there. When $s$ is small, we concentrate on the poles and singularities near zero. There might be other singularities that stay at a fixed $z$ when we take $s \to0$. We do not indicate those here since they are further away in the $z$ plane.  
     (b) After a contour deformation, we pick up contributions from the (red) poles $z_k$ in \nref{zkpole} as well as some contributions along the branch cuts (in blue). Finally, we have a contribution from a circle whose size we keep fixed as we take $s\to 0$.    }
    \label{Contour}
 \end{figure}

We now write 
\be 
{\cal A}_{n+1} = { 1 \over 2\pi i } \oint { d z \over z } {\cal A}_{n+1}(z) 
\ee 
where the integral is over a very small circle around the origin which only encloses the explicit pole at $z=0$, see figure \ref{Contour}.

As a function of $z$, ${\cal A}(z)$ has some poles and branch cuts, see figure \ref{Contour}. We will now deform the contour in such a way that we pick up the contributions of the poles that are close to the origin. In doing this, there might also be contributions from branch cuts and the remaining contour integral at some finite value of $|z|$. In contrast to \cite{Britto:2005fq,Arkani-Hamed:2008owk}, we will not make any assumptions about the behavior of the amplitude for very large $z$ because we keep the contour at a finite distance. 

The poles that are near the origin are the ones that come from $(p_k +q)^2 \to 0$. These poles sit at 
\be \la{zkpole}
\langle s(z) , k\rangle =0 = \langle s , k \rangle + z \langle n, k \rangle ~~~~~\longrightarrow  ~~~~~~~z_k \equiv - { \langle s,k \rangle 
\over \langle n, k\rangle } 
\ee 
where, as usual $\langle a, b \rangle = \epsilon^{\alpha \beta } \lambda^a_\alpha \lambda^b_\beta $. As expected, $z_{k}$ approaches the origin in the soft limit, $|s\rangle \to 0$. The behavior of the amplitude at this pole is determined by our factorization assumption \nref{FacPo}:
\be\label{bcfwt}
\textrm{Res}_{z=z_{k}}\left (\frac{\mathcal{A}_{n+1}}{z}\right )\propto 
\frac{\mathcal{A}_{3}(z_{k})\mathcal{A}_{n}(z_{k})}{[s,k] \langle s, k \rangle } 
\ee
 where $\mathcal{A}_{3}(z_{k})$ and $\mathcal{A}_{n}(z_{k})$ correspond to the three and $n$-point amplitudes with deformed, but still on-shell, kinematics.  To evaluate the residue at the pole \nref{zkpole} we need the spinor helicity variables for the intermediate particle, $I$:
 \be \la{IntVa}
p_I = |I \rangle [I |=|s\rangle [s|+|k\rangle[k|+z_{k}|n\rangle[s| ~,~~~~~~~~ |  I \rangle = |k \rangle ~,~~~~~~~~|I] = |k] + { \langle n , s \rangle \over \langle n , k \rangle } |s]   \ .
 \ee 
We also need the on-shell three-point amplitude with momenta $q$, $p_k$, and $p_I$. We assume that the soft particle has positive helicity, and helicities $h$ and $-h$ for particles $k$ and $I$, which correspond to minimal coupling\footnote{As a side comment, there are three-point amplitudes not of this form, such as the ones coming from an  $R^{3}$ contribution,  but they involve non-minimal couplings in the gravitational theory and are not present in the matrix model three point amplitude. According to  \cite{Elvang:2016qvq}, no local coupling can impact the leading and subleading soft graviton theorems.}. The three-point amplitude is then  
 \be \la{Ampl}
 {\cal A}_{3}(z_{k}) \propto  [ s,k]^{2 + 2 h } [ s, I]^{ 2 - 2 h}   [k,I]^{ - 2 } \to [ s, k]^2 { \langle n,k\rangle^2 \over \langle n, s \rangle^2 } 
 \ee 
 where in the last expression we used \nref{IntVa}. 
We have ignored the overall factor of $\sqrt{G_N}$ and other numerical factors. The ${\cal A}_{n}(z_{k})$ amplitude can also be written as a function of external kinematics: 
\bea\label{Ampr}
&~& ~~~~~~~\mathcal{A}_{n}(1, \cdots ,   I(z) , \cdots ,   n(z) ) ~,~~~~~~{\rm with}
\cr 
 &~& ~|I(z)] =   |k]+\frac{\langle n,s\rangle }{\langle n,k\rangle }|s],~,~~~~~|I\rangle = |k\rangle~,~~~~~~~|n(z_{k})]= |n]+\frac{\langle s,k\rangle}{\langle n,k\rangle}|s] \ .
\eea
where the intermediate momentum appears in the $k^{th}$ position. Note that (\ref{Ampl}) and (\ref{Ampr}) are exact and only the square brackets are modified. We have not yet taken the soft limit.

 Using \nref{bcfwt}, \nref{Ampl} and \nref{Ampr}, we can determine the residue on the pole. After summing over all residues, the expression for the amplitude becomes
 \be \la{PolCo} 
 {\cal A}_{\rm poles } \propto \sum_{k \not =n} { \langle n , k \rangle^2 \over \langle  n,s \rangle^2 } { [ s , k ] \over \langle s, k \rangle }   \mathcal{A}_n\left( |1] , \cdots,    |k]+\frac{\langle n,s\rangle}{\langle n,k\rangle}|s], \cdots, |n] + \frac{\langle s,k\rangle}{\langle n,k\rangle}|s]  \right ) \ .
 \ee 
 The amplitude $\mathcal{A}_n$ contains $n$ particles which are the original $n+1$ particles with the soft one removed. In the arguments of ${\cal A}_n$,  the angle brackets are unshifted, but two of the square brackets are shifted as indicated. 
  The leading soft factor comes from setting $s=0$ in the ${\cal A}_n$ amplitude in \nref{PolCo}, so that it is evaluated in the original variables,       
  \be \la{SoftL}
   S_0 {\cal A}_n \propto  \sum_{k  } { \langle n , k \rangle^2 \over \langle  n,s \rangle^2 } { [ s , k ] \over \langle s, k \rangle } {\cal A}_n   
  \ee 
   where we also extended the sum to include $k=n$ since this term vanishes anyway.   The amplitude ${\cal A}_n$ is evaluated on the momenta $p_1, \cdots, p_n$, the undeformed momenta, which conserve momentum when $s=0$      
     
  The subleading soft factor comes from expanding \nref{PolCo} to the next order. 
  When performing this expansion, it is convenient to keep $|s]$ fixed and expand in $|s\rangle$. 
  The next term in the expansion is then
 \bea \la{DsFirst}
 S_1 {\cal A}_n = &=&\half \sum_{k \not = n } { \langle n,k\rangle \over \langle n,s \rangle^2  } { [ s, k ] \over \langle s,k \rangle} \bar \lambda^s_{\dot \beta } \left \{ \langle n, s\rangle  { \partial {\cal A}_n \over \partial \bar \lambda^k_{\dot \beta }  }  + \langle k, s \rangle {\partial {\cal A}_n \over \partial   \bar \lambda^n_{\dot \beta } }  \right\} 
 \\ 
  & = & \half \sum_{k  } { \langle n,k\rangle \over \langle n,s \rangle   } { [ s, k ] \over \langle s,k \rangle}  \bar \lambda^s_{\dot \beta }  { \partial {\cal A}_n \over \partial \bar \lambda^k_{\dot \beta }  }  \la{SinfdS}  
  \eea 
 where, in the second line, we used momentum conservation   
 \be 
  \sum_k | k \rangle |k ] = - |s \rangle |s] - |n \rangle | n ] 
  \ee 
  to get rid of the second term in \nref{DsFirst}. 
  
  So far, we have argued that the pole contributions give us what we want. We will now argue that nothing else contributes. There are two sources of potentially undesirable contributions
  \begin{itemize}
\item The contour deformation argument gives us contributions from the branch cuts. 
  Our assumption 3) is that these contributions scale like a positive power of $q$ and that we do not need to consider it at leading and subleading order in the soft limit.
  \item There is also a contribution coming from the amplitude evaluated on a circle of small but fixed size as $|s\rangle \to 0$.
   In this integral, one would think that we could simply set $|s\rangle \to 0$. In that case, the soft momentum and the $n^{th}$ momentum go as
   \be \la{Psn}
   p_s = z |n\rangle |s ] ~,~~~~~~~~p_n = |n\rangle \left(  |n] - z|s] \right)
   \ee 
   in the soft limit. Crucially, both momenta are finite. Unfortunately, we have that $p_{I}^{2}=(p_{s}+p_{n})^{2} \to 0$, which would lead to a divergent contribution in the soft limit. Note that $p_I$ is finite but null in this limit.  Therefore, we need to examine more closely the approach to $|s \rangle \to 0$. We use again the factorization property \nref{FacPo}. The denominator vanishes as $\langle s | n \rangle$. The three-point amplitude and the $n$-point amplitude are both finite as $|s\rangle \to 0$ because none of the relevant momenta, \nref{Psn}, $p_{I}$ and the other external momenta, are soft for finite values of $z$. This means that this term goes as $1/s$. This is subleading compared to 
   \nref{SoftL} and \nref{SinfdS}, which go like $1/s^3$ and $1/s^2$ respectively. Since this term is subleading in $s$ for any value of $z$ on the outer contour in figure \ref{Contour}, the outer contour does not contribute to the leading and subleading soft limits. 
   In principle that is all we need to say. However, we can also notice that when we factor on this pole the intermediate momentum becomes just $p_I = p_n$ and the amplitude ${\cal A}_n(p_1 , \cdots, p_I)$ becomes independent of $z$. The $z$ dependence of the three-point amplitude is easy to calculate. The three square brackets of the three particles are $|s], ~|n] - z |s], ~|n] $  and the on shell three point amplitudes goes as ${\cal A}_3 \propto z^{-2}$. Then, the final expression goes as $ \oint d z z^{-3}  = 0 $. In other words, the $z$ integral of the term that is going like ${ 1 \over \langle s , n\rangle}$ vanishes. 
  \end{itemize}
  Therefore, we expect that the remaining $z$ integral is finite as $|s\rangle \to 0$. This concludes our argument for the leading and subleading soft factors \nref{SoftL} and \nref{SinfdS}. It looks like we can also write an expression for the subsubleading soft factor by expanding \nref{PolCo} to one higher order. We have not done so because it will not be necessary for what we want to do in this paper.

  \subsection{A contour deformation argument involving a three line shift} 
  
  The reader might be slightly confused by our treatment of the part of the integral that does not involve the poles. We can present an argument that is conceptually a bit clearer by performing a different shift of the spinor helicity variables when we go to the complex plane. Instead of the BCFW shift, we instead consider the three-line shift
  \bea 
  |s \rangle &\to &  |s(z) \rangle = |s\rangle + z [1,2] |\eta\rangle 
  \cr 
   |1 \rangle &\to &  |1(z) \rangle = |1\rangle + z [2,s] |\eta\rangle 
\cr 
|2 \rangle &\to &  |2(z) \rangle = |2\rangle + z [s,1] |\eta\rangle 
  \eea
  where $|\eta\rangle$ is an arbitrary vector \cite{Risager:2005vk,Bjerrum-Bohr:2005xoa}. We again take the soft limit by sending $|s\rangle \to 0$ keeping $|s]$ fixed. The primary advantage of the three-line shift is that when we set $|s\rangle =0$, all momenta are finite for general $z$ and we are generically away from any singularity. Therefore, after we deform the contour, we can then set $|s\rangle =0$ and this integral is manifestly finite.\footnote{Amusingly, this three-line shift actually fails when used to re-construct tree-level gravity amplitudes starting at 12-point due to a boundary term \cite{Bianchi:2008pu,Cohen:2010mi}. This boundary term is irrelevant for our discussion because we are not taking the contour to infinity.} 
  
  Of course, when we deform the contour, we pick up contributions from the poles. Again, the only poles at small values of $z$ come from those that diverge in the soft limit. The contributions from these poles can be analyzed and expanded as above. Repeating similar algebraic steps to those above, we ultimately get the leading and subleading soft factors 
 \be \la{3Llea}
 S_0 {\cal A}_n \propto \sum_k   { \langle k , \eta\rangle ^2 \over \langle s, \eta \rangle^2 } { [ s, k ] \over \langle s , k \rangle }  {\cal A}_n 
 \ee 
 \be \la{3Lslea}
  S_1 {\cal A}_n \propto \sum_k { \langle k , \eta\rangle   \over \langle s, \eta \rangle  } { [ s, k ] \over \langle s , k \rangle } \bar \lambda^s_{\dot \beta} { \partial {\cal A}_n \over \partial \bar \lambda^k_{\dot \beta } } \ .
  \ee 
  These are similar to \nref{PolCo} and \nref{SinfdS} but with $|n\rangle \to |\eta \rangle  $ in the prefactors.

  \subsection{Momentum conservation and Lorentz Symmetry } 
  
  The expressions derived above for the soft factors involve some arbitrary vectors. In \nref{3Llea} and \nref{3Lslea}, there is an auxiliary spinor $|\eta\rangle$. In \nref{SoftL} and \nref{SinfdS}, the auxiliary vector is replaced by a spinor of the particle we used for the BCFW shift, which was particle $n$ in section \ref{BCFWsec}. 
  
  In both of these cases, these special vectors appeared as some arbitrary choices when we performed a complex deformation of the amplitude. However, the amplitude itself is independent of these choices. Therefore, as a consistency condition, we should impose that \nref{3Llea} and \nref{3Lslea} are independent of $|\eta\rangle$. 
  Notice that the expressions are already independent of the overall scaling of $|\eta\rangle$. In addition, we can demand that they are invariant under $|\eta \rangle \to |\eta \rangle + \epsilon |s\rangle$. This shift cancels the $\langle s, k \rangle$ in the denominators and \nref{3Llea} yields the constraint
  \be 
  0 = { 1 \over \langle s, \eta \rangle^2 } \sum_k  \langle k , \eta \rangle [ s, k ] { \cal A}_n 
  \ee 
  which is momentum conservation. Similarly, equation \nref{3Lslea} yields
  \be 
   0 = { 1 \over \langle s, \eta \rangle  } \sum_k    \bar \lambda^s_{\dot \alpha } \bar \lambda^s_{\dot \beta} \bar \lambda^{k \dot \alpha } { \partial {\cal A}_n \over \partial \bar \lambda^k_{\dot \beta } } ~~\to ~~  0=\bar \lambda^s_{\dot \alpha } \bar \lambda^s_{\dot \beta} \sum_k    { \cal L}_k^{\dot \alpha \dot \beta }  {  {\cal A}_n   }
  \ee 
 which is the condition for invariance under the $SL(2)$ symmetry acting on the $\bar \lambda^k_{\dot \beta}$ variables, where 
 \be 
 {\cal L}_k^{\dot \alpha \dot \beta } =  \lambda^{k \dot \alpha } { \partial  \over \partial \bar \lambda^k_{\dot \beta } } +  \lambda^{k \dot \beta } { \partial  \over \partial \bar \lambda^k_{\dot \alpha } } \ . 
  \ee 
 If we had considered a {\it negative} helicity soft graviton, we would have obtained the $SL(2)$ that acts on the $\lambda^k_\alpha$ variables. Of course, we are using that, suitably complexified, $SO(1,3) \sim SL(2) \times SL(2)$.

We could have obtained the same conclusions from \nref{SoftL} and \nref{SinfdS}. To see this, note that the answer should be independent of the particle we pick to do the BCFW shift. If we had chosen a particle $m$ instead of particle $n$, we would then have obtained \nref{SoftL} and \nref{SinfdS} with $|n\rangle \to | m\rangle$. For each external $m\neq n$, we set to zero the difference between the expression with $|n\rangle$ and the one with $|m\rangle$, again finding the momentum conservation and Lorentz invariance conditions. 
 
 The conclusion is that consistency of the soft theorem implies the Lorentz symmetry of the amplitude, as emphasized in \cite{Tropper:2023fjr}. This fact can be physically understood as follows. The soft graviton represents the coupling to long-distance gravity, and only theories that are Poincare invariant in flat space can be consistently coupled to gravity. 
 
 \section{Application to the BFSS matrix model } \label{bfssmod}

 \subsection{Lightning review of the BFSS conjecture }

 In the BFSS matrix model, one considers a scattering problem in a matrix quantum mechanics of size $N$, which has a $U(N)$ gauge symmetry. 
  The model has nine matrices and a classical potential that has flat directions involving diagonal matrices. 
  The model has degrees of freedom in its $U(1)$ and $SU(N)$ sectors. The $SU(N)$ sector contains a zero-energy bound state \cite{Yi:1997eg,Sethi:1997pa,Moore:1998et,Konechny:1998vc,Porrati:1997ej,Sethi:2000zf,Lin:2014wka}. On top of this bound state,  the $U(1)$ part describes the degrees of freedom of a massless superparticle in eleven dimensions in light-cone gauge where $p_- =- N/R_-$,  which describes the graviton supermultiplet.  
  
  The model also has some asymptotic directions where the matrix splits into blocks of matrices of sizes $N_i$, $U(N) \to \prod_i U(N_i)$. Then, in each $SU(N_i)$, we have the zero energy ground state and the overall $U(1)_i$ degrees of freedom in $U(N_i)$ describe a massless superparticle. The incoming and outgoing states could have different splittings of $N = \sum_i N_i$. The conjecture is that the scattering amplitude in the quantum mechanical model describes the scattering amplitude in M-theory. 
  More precisely, the idea is that in the large $N$ limit, where 
  \be \label{bfsslm}
  N_i \to \infty ~,~~~~~R_- \to \infty ~,~~~~~~~~ -p_-^i = { N_i \over R_- } = {\rm fixed} \ ,
  \ee 
  the scattering amplitude in the matrix model is equal to the one in the eleven-dimensional M-theory. Each particle also involves a   momentum vector $\vec p^{\, \,i}$ which arises due to the momentum of the $i^{th}$ sub-block in the nine transverse dimensions. Notice that $p_+^{i}$ is fixed by the on-shell condition for a massless particle. 
  
 The BFSS conjecture implies several sub-conjectures in the limit \nref{bfsslm}:
 \begin{itemize}
 	\item[I:]  
 The limit \nref{bfsslm} is finite and defines a continuous and suitably analytic function of the $p_-^i$ and $\vec p^{\, \, i}$.  
 
 \item[II:] The limit defines a function that is Lorentz invariant in eleven dimensions.  
    
    \item[III:] The limit has all other desired physical properties of eleven-dimensional M-theory. 
   
  \end{itemize}
  In this paper, we assume I,  which is the same as 1), in section \ref{Assumptions}. We will then argue that all other assumptions in that section are true so that II follows. 
  We do not say anything about III in this paper.  
  
  \subsection{Properties of the matrix model } 
  
  Here we argue for properties 2) and 3) in section \ref{Assumptions}. Property 4), which is the usual form for the three point function, was demonstrated in \cite{Herderschee:2023pza}. 
  
 It is important to understand the origin of singularities in the matrix model amplitudes. 
  For finite $N$, the matrix model problem is a scattering problem in quantum mechanics. 
  If we use old-fashioned perturbation theory, a singularity can arise when there are energy denominators that are becoming zero. For example, we can have a singularity that arises because two sub-blocks merge, or are close to merging into some intermediate state. In this region, we are talking about an energy denominator of the form 
  \be \la{EnDen}
  \sum_I  { 1 \over E_I - E_{ext} } {\cal M} 
  \ee 
  where we sum over intermediate states of energies $E_I$. The external energies are the energies of the particles $s$ and $k$ in the quantum mechanical model. And ${\cal M}$ denotes some matrix elements, which we assume are smooth in the region we consider. 
  Here, we are considering a block of size $N_I$ that is splitting into two subblocks of size $N_s$ and $N_k$. Each of these particles obeys an on-shell condition 
  \be \la{OnShec}
  2 p_- p_+ = \vec p^{\,\, 2} ~,~~~~~~~~ - p_- = { N \over R_-} \ .
  \ee 
  We then see that $E_{ext} = - p_+^s - p_+^k $ and it can be rewritten using \nref{OnShec}. 
  
  The intermediate state $I$ lives in the $U(N_I)$ subsector. The overall momentum in the  $U(1)_I$, i.e. the center of mass sector, is fixed to be $\vec p_I = \vec p_s + \vec p_k$. 
  We can then write the energy of the intermediate state as 
  \be 
  E_I = \half { \vec p_I^{\,\, 2} R_- \over  N_I }  + \epsilon_I 
  \ee 
  where the first term comes from the $U(1)_I$ sector and $\epsilon_I$ is the energy in the $SU(N_I)$ subsector. We have $\epsilon_I =0$ for the zero energy intermediate state. 
  Using \nref{OnShec} for the external particles, we can rewrite 
  \be  \la{DeH}
  E_I - E_{ext} = { p^s \cdot p^k \over   N_I } + \epsilon_I ~,~~~~~~~p^s\cdot p^k = p^s_+ {N_k \over R_-} + p^k_+ {N_s \over R_-} + \vec p^{\, s} \cdot \vec p^{\, k} \ .
  \ee 
  We see that the bound state with $\epsilon_I =0$ gives rise to the pole discussed in 
  \nref{FacPo}. In that case, the matrix element in \nref{EnDen} involves a three-point amplitude and an amplitude with one less particle. This result follows from factorization of the matrix model scattering amplitude on an on-shell intermediate state, which we would have at the pole $E_I - E_{ext}=0$, when $\epsilon_I =0$. 
  Note that this denominator can only vanish when we analytically continue in the external momenta; it is always bigger than zero for physical incoming momenta. 
  
  We conclude that the $SU(N_I)$ bound state gives rise to the pole and factorization structure in \nref{FacPo}. The odd-looking factor of $N_I$ in \nref{DeH}  arises when we compare the relativistic and non-relativistic normalizations, see \cite{Herderschee:2023pza}. 
  
  We need to take into account the continuum of states that exist above zero energy, $\epsilon_I > 0$ \cite{deWit:1988xki}. This continuum arises due to the existence of a large set of scattering states. Here we will consider the simplest one. We imagine that the $N_I$ block splits into two blocks of sizes $N_1$ and $N_2$. The center of mass degree of freedom, the $U(1)$ in $U(N_I)$ is constrained as before. The relative degree of freedom arises from a combination of the $U(1)$ in $U(N_1)$ and the $U(1)$ in $U(N_2)$. It gives us the relative position of these two blocks. The energy takes the form 
  \be 
  \epsilon_I = \half \vec p_r^{\, 2} { R_- \over N_r} ~,~~~~~~ \vec p_r \equiv { N_2 \vec p_1 - N_1  \vec p_2 \over N_I }  ~,~~~~~~~~~~ N_r \equiv  {N_1 N_2 \over N_I } ~,~~~~N_I = N_1 + N_2 
  \ee 
  Integrating over $\vec p_r$ yields an expression proportional to 
  \be \la{INtbc}
  \int d^9 p_r \, \, { 1 \over   \half \vec p_r^{\, 2 } + { N_r \over N_I } p^s \cdot  p^k } 
  \ee 
   The crucial feature of this integral is that it is finite in the IR, where  $\vec p_r \sim  0$, thanks to the fact that we are in a large number of dimensions. If we think of $ \zeta = p^s \cdot p^k$ as a complex variable, we get a branch cut at $\zeta=0$ of the form $\zeta^{7/2}$. If we had done the same in $d$ transverse dimensions, we would get $\zeta^{d-2\over 2}$.
   
   This branch cut involves a large power of $\zeta$, so that its contribution as $p_s \to 0$ is suppressed relative to the terms we kept. We expect that states with larger numbers of particles will give rise to branch cuts that are similarly suppressed in the soft limit. 
   
   Notice that in the argument in section \ref{GenSoft}, the integral \nref{INtbc} gives rise to a branch cut in ${\cal A}_{n+1}(z)$ which starts at $z= z_k$ and goes like $(z-z_k)^{7/2}$, see figure \ref{Contour}.  The integral along this branch cut of the original expression has the form 
   \be \la{BrCon}
   \int { d z \over z } ( z - z_k)^{7/2} 
   \ee 
   which has a finite limit when $z_k \to 0$, which is the limit when $|s\rangle \to 0$. This branch cut is also expected in ordinary gravity amplitudes when we go to higher loops. Therefore the soft argument of section \nref{GenSoft} also implies that higher loop amplitudes obey the soft theorem in high enough total number of dimensions, namely  $D> 4$ \cite{Bern:2014oka,Bern:2014vva,He:2014bga,Broedel:2014fsa,Sen:2017nim,Krishna:2023fxg}.

Note that our arguments for the pole and branch cut were for finite $N$. In principle, the large $N$ limit might produce new singularities. We are assuming that any new singularities, if present, will give a subleading contribution when $q_s \to 0$, similarly to the branch cuts we explicitly discussed\footnote{In the large $N$ limit, new singularities appear in $AdS/CFT$ examples, where they are associated with the emergence of the bulk \cite{Gary:2009ae,Maldacena:2015iua}. In fact, we expect such new singularities in the BFSS model when we consider it at higher energies,  where it has a holographic description in terms of a ten-dimensional black hole \cite{Itzhaki:1998dd}.  Numerics can probably be applied to this question along the lines of \cite{Bergner:2021goh}.}.

   Note that in the BFSS model, we can use the explicit symmetry of the matrix model to transform any four-point amplitude into one where the kinematics is four-dimensional. This can be done as follows. First, consider the two incoming particles and use a Galilean transformation to set $\vec p_1 + \vec p_2 =0$. Then use a rotation to put these two vectors along the first of the nine dimensions. By a similar rotation,  the momenta of the two outgoing particles can be set along the first and second spatial transverse dimensions. Together with the two light-cone directions, these vectors lie in a four-dimensional subspace. Therefore, our arguments show that the four-point amplitude is four-dimensional Lorentz invariant, which in turn, using the explicit symmetries of the matrix model, implies the eleven-dimensional Lorentz symmetry.   
   
   On the other hand,  for $n \geq 5$, our arguments only show Lorentz invariance for amplitudes whose momenta lie in a four-dimensional subspace. One would like to have a more general argument that is valid for general eleven-dimensional kinematics. We leave this problem to the future.

    \section{Conclusions } \label{conc}
    
    The goal of this article was to show that the BFSS model amplitudes obey the gravitational soft theorems. This in turn implies that the amplitudes are Lorentz invariant. As an input, we assumed that the eleven-dimensional limit \nref{bfsslm} is well defined. We used the factorization properties of the amplitude and the explicit form of the three-point amplitude. The argument used a BCFW deformation of the momenta. 

    Note that in our arguments, we took the soft momentum to have non-zero longitudinal momentum $N_s$, and took the $N_s \to \infty $ limit to obtain a continuous function of $p^s_- = -N_s/R_-$  before taking the soft limit. We then took $|p^s_-|$ to be much smaller than the other $|p_-^k|$.  The nonzero value for $q_- = p^s_-$ is crucial for extracting, from the subleading soft theorem, the Lorentz generators that mix the $x^-$ direction with the others. These Lorentz generators contain a derivative with respect to $p_-$ and such terms arise in the subleading soft theorem only if $q_-$ is non-zero.

     BCFW shifts have been used to derive the full gravity tree amplitude starting from the three-point amplitude \cite{Cachazo:2005ca,Bedford:2005yy,Hodges:2011wm}.  It would be nice to use a similar BCFW shift in the matrix model. One might need to similarly argue that the branch-cut contributions are subleading at low energies and that the contour vanishes at infinity.

 Our analysis was restricted to a four-dimensional kinematic configuration because spinor helicity variables are convenient. This analysis is sufficient to derive the Lorentz symmetry of the four-point amplitude, but it is not general enough for higher point amplitudes. Extending the argument to a general kinematic configuration appears to be a matter of relatively simple algebra and the same basic strategy should work.    
          
\subsection*{Acknowledgments}

We thank T. Banks, N. Miller, S. Mizera, N. Seiberg, A. Strominger, A. Tropper, N. Valdes-Meller, H. Verlinde, and T. Wang for discussions.   

J.M. is supported in part by U.S. Department of Energy grant DE-SC0009988. A.H. is supported by the Simons Foundation. 

 \bibliographystyle{apsrev4-1long}
\bibliography{GeneralBibliography.bib}

\begin{thebibliography}{10}%
\makeatletter
\providecommand \@ifxundefined [1]{%
 \ifx #1\undefined \expandafter \@firstoftwo
 \else \expandafter \@secondoftwo
\fi
}%
\providecommand \@ifnum [1]{%
 \ifnum #1\expandafter \@firstoftwo
 \else \expandafter \@secondoftwo
\fi
}%
\providecommand \enquote [1]{``#1''}%
\providecommand \bibnamefont  [1]{#1}%
\providecommand \bibfnamefont [1]{#1}%
\providecommand \citenamefont [1]{#1}%
\providecommand\href[0]{\@sanitize\@href}%
\providecommand\@href[1]{\endgroup\@@startlink{#1}\endgroup\@@href}%
\providecommand\@@href[1]{#1\@@endlink}%
\providecommand \@sanitize [0]{\begingroup\catcode`\&12\catcode`\#12\relax}%
\@ifxundefined \pdfoutput {\@firstoftwo}{%
 \@ifnum{\z@=\pdfoutput}{\@firstoftwo}{\@secondoftwo}%
}{%
 \providecommand\@@startlink[1]{\leavevmode\special{html:<a href="#1">}}%
 \providecommand\@@endlink[0]{\special{html:</a>}}%
}{%
 \providecommand\@@startlink[1]{%
  \leavevmode
  \pdfstartlink
   attr{/Border[0 0 1 ]/H/I/C[0 1 1]}%
   user{/Subtype/Link/A<</Type/Action/S/URI/URI(#1)>>}%
  \relax
 }%
 \providecommand\@@endlink[0]{\pdfendlink}%
}%
\providecommand \url  [0]{\begingroup\@sanitize \@url }%
\providecommand \@url [1]{\endgroup\@href {#1}{\urlprefix}}%
\providecommand \urlprefix [0]{URL }%
\providecommand \Eprint[0]{\href }%
\@ifxundefined \urlstyle {%
  \providecommand \doi [1]{doi:\discretionary{}{}{}#1}%
}{%
  \providecommand \doi [0]{doi:\discretionary{}{}{}\begingroup
  \urlstyle{rm}\Url }%
}%
\providecommand \doibase [0]{http://dx.doi.org/}%
\providecommand \Doi[1]{\href{\doibase#1}}%
\providecommand \bibAnnote [3]{%
  \BibitemShut{#1}%
  \begin{quotation}\noindent
    \textsc{Key:}\ #2\\\textsc{Annotation:}\ #3%
  \end{quotation}%
}%
\providecommand \bibAnnoteFile [2]{%
  \IfFileExists{#2}{\bibAnnote {#1} {#2} {\input{#2}}}{}%
}%
\providecommand \typeout [0]{\immediate \write \m@ne }%
\providecommand \selectlanguage [0]{\@gobble}%
\providecommand \bibinfo [0]{\@secondoftwo}%
\providecommand \bibfield [0]{\@secondoftwo}%
\providecommand \translation [1]{[#1]}%
\providecommand \BibitemOpen[0]{}%
\providecommand \bibitemStop [0]{}%
\providecommand \bibitemNoStop [0]{.\EOS\space}%
\providecommand \EOS [0]{\spacefactor3000\relax}%
\providecommand \BibitemShut [1]{\csname bibitem#1\endcsname}%
\bibitem{deWit:1988wri}%
  \BibitemOpen
  \bibfield{author}{%
  \bibinfo {author} {\bibfnamefont{B.}~\bibnamefont{de~Wit}}, \bibinfo {author}
  {\bibfnamefont{J.}~\bibnamefont{Hoppe}},\ and\ \bibinfo {author}
  {\bibfnamefont{H.}~\bibnamefont{Nicolai}},\ }%
  \bibfield{title}{%
  \enquote{\bibinfo {title} {{On the Quantum Mechanics of Supermembranes}},}\
  }%
  \bibfield{journal}{%
  \Doi{10.1016/0550-3213(88)90116-2}{\bibinfo {journal} {Nucl. Phys. B}}\ }%
  \textbf{\bibinfo {volume} {305}},\ \bibinfo {pages} {545} (\bibinfo {year}
  {1988})%
  \bibAnnoteFile{NoStop}{deWit:1988wri}%
\bibitem{Banks:1996vh}%
  \BibitemOpen
  \bibfield{author}{%
  \bibinfo {author} {\bibfnamefont{Tom}\ \bibnamefont{Banks}}, \bibinfo
  {author} {\bibfnamefont{W.}~\bibnamefont{Fischler}}, \bibinfo {author}
  {\bibfnamefont{S.~H.}\ \bibnamefont{Shenker}},\ and\ \bibinfo {author}
  {\bibfnamefont{Leonard}\ \bibnamefont{Susskind}},\ }%
  \bibfield{title}{%
  \enquote{\bibinfo {title} {{M theory as a matrix model: A Conjecture}},}\ }%
  \bibfield{journal}{%
  \Doi{10.1103/PhysRevD.55.5112}{\bibinfo {journal} {Phys. Rev. D}}\ }%
  \textbf{\bibinfo {volume} {55}},\ \bibinfo {pages} {5112--5128} (\bibinfo
  {year} {1997}),\
  \Eprint{http://arxiv.org/abs/hep-th/9610043}{arXiv:hep-th/9610043}%
  \bibAnnoteFile{NoStop}{Banks:1996vh}%
\bibitem{Susskind:1997cw}%
  \BibitemOpen
  \bibfield{author}{%
  \bibinfo {author} {\bibfnamefont{Leonard}\ \bibnamefont{Susskind}},\ }%
  \bibfield{title}{%
  \enquote{\bibinfo {title} {{Another conjecture about M(atrix) theory}},}\ }%
   (\bibinfo {month} {4}\ \bibinfo {year} {1997}),\
  \Eprint{http://arxiv.org/abs/hep-th/9704080}{arXiv:hep-th/9704080}%
  \bibAnnoteFile{NoStop}{Susskind:1997cw}%
\bibitem{Seiberg:1997ad}%
  \BibitemOpen
  \bibfield{author}{%
  \bibinfo {author} {\bibfnamefont{Nathan}\ \bibnamefont{Seiberg}},\ }%
  \bibfield{title}{%
  \enquote{\bibinfo {title} {{Why is the matrix model correct?}}.}\ }%
  \bibfield{journal}{%
  \Doi{10.1103/PhysRevLett.79.3577}{\bibinfo {journal} {Phys. Rev. Lett.}}\ }%
  \textbf{\bibinfo {volume} {79}},\ \bibinfo {pages} {3577--3580} (\bibinfo
  {year} {1997}),\
  \Eprint{http://arxiv.org/abs/hep-th/9710009}{arXiv:hep-th/9710009}%
  \bibAnnoteFile{NoStop}{Seiberg:1997ad}%
\bibitem{Sen:1997we}%
  \BibitemOpen
  \bibfield{author}{%
  \bibinfo {author} {\bibfnamefont{Ashoke}\ \bibnamefont{Sen}},\ }%
  \bibfield{title}{%
  \enquote{\bibinfo {title} {{D0-branes on T**n and matrix theory}},}\ }%
  \bibfield{journal}{%
  \Doi{10.4310/ATMP.1998.v2.n1.a2}{\bibinfo {journal} {Adv. Theor. Math.
  Phys.}}\ }%
  \textbf{\bibinfo {volume} {2}},\ \bibinfo {pages} {51--59} (\bibinfo {year}
  {1998}),\ \Eprint{http://arxiv.org/abs/hep-th/9709220}{arXiv:hep-th/9709220}%
  \bibAnnoteFile{NoStop}{Sen:1997we}%
\bibitem{Polchinski:1999br}%
  \BibitemOpen
  \bibfield{author}{%
  \bibinfo {author} {\bibfnamefont{Joseph}\ \bibnamefont{Polchinski}},\ }%
  \bibfield{title}{%
  \enquote{\bibinfo {title} {{M theory and the light cone}},}\ }%
  \bibfield{journal}{%
  \Doi{10.1143/PTPS.134.158}{\bibinfo {journal} {Prog. Theor. Phys. Suppl.}}\
  }%
  \textbf{\bibinfo {volume} {134}},\ \bibinfo {pages} {158--170} (\bibinfo
  {year} {1999}),\
  \Eprint{http://arxiv.org/abs/hep-th/9903165}{arXiv:hep-th/9903165}%
  \bibAnnoteFile{NoStop}{Polchinski:1999br}%
\bibitem{Yi:1997eg}%
  \BibitemOpen
  \bibfield{author}{%
  \bibinfo {author} {\bibfnamefont{Piljin}\ \bibnamefont{Yi}},\ }%
  \bibfield{title}{%
  \enquote{\bibinfo {title} {{Witten index and threshold bound states of
  D-branes}},}\ }%
  \bibfield{journal}{%
  \Doi{10.1016/S0550-3213(97)00486-0}{\bibinfo {journal} {Nucl. Phys. B}}\ }%
  \textbf{\bibinfo {volume} {505}},\ \bibinfo {pages} {307--318} (\bibinfo
  {year} {1997}),\
  \Eprint{http://arxiv.org/abs/hep-th/9704098}{arXiv:hep-th/9704098}%
  \bibAnnoteFile{NoStop}{Yi:1997eg}%
\bibitem{Sethi:1997pa}%
  \BibitemOpen
  \bibfield{author}{%
  \bibinfo {author} {\bibfnamefont{Savdeep}\ \bibnamefont{Sethi}}\ and\
  \bibinfo {author} {\bibfnamefont{Mark}\ \bibnamefont{Stern}},\ }%
  \bibfield{title}{%
  \enquote{\bibinfo {title} {{D-brane bound states redux}},}\ }%
  \bibfield{journal}{%
  \Doi{10.1007/s002200050374}{\bibinfo {journal} {Commun. Math. Phys.}}\ }%
  \textbf{\bibinfo {volume} {194}},\ \bibinfo {pages} {675--705} (\bibinfo
  {year} {1998}),\
  \Eprint{http://arxiv.org/abs/hep-th/9705046}{arXiv:hep-th/9705046}%
  \bibAnnoteFile{NoStop}{Sethi:1997pa}%
\bibitem{Moore:1998et}%
  \BibitemOpen
  \bibfield{author}{%
  \bibinfo {author} {\bibfnamefont{Gregory~W.}\ \bibnamefont{Moore}}, \bibinfo
  {author} {\bibfnamefont{Nikita}\ \bibnamefont{Nekrasov}},\ and\ \bibinfo
  {author} {\bibfnamefont{Samson}\ \bibnamefont{Shatashvili}},\ }%
  \bibfield{title}{%
  \enquote{\bibinfo {title} {{D particle bound states and generalized
  instantons}},}\ }%
  \bibfield{journal}{%
  \Doi{10.1007/s002200050016}{\bibinfo {journal} {Commun. Math. Phys.}}\ }%
  \textbf{\bibinfo {volume} {209}},\ \bibinfo {pages} {77--95} (\bibinfo {year}
  {2000}),\ \Eprint{http://arxiv.org/abs/hep-th/9803265}{arXiv:hep-th/9803265}%
  \bibAnnoteFile{NoStop}{Moore:1998et}%
\bibitem{Konechny:1998vc}%
  \BibitemOpen
  \bibfield{author}{%
  \bibinfo {author} {\bibfnamefont{Anatoly}\ \bibnamefont{Konechny}},\ }%
  \bibfield{title}{%
  \enquote{\bibinfo {title} {{On asymptotic Hamiltonian for SU(N) matrix
  theory}},}\ }%
  \bibfield{journal}{%
  \Doi{10.1088/1126-6708/1998/10/018}{\bibinfo {journal} {JHEP}}\ }%
  \textbf{\bibinfo {volume} {10}},\ \bibinfo {pages} {018} (\bibinfo {year}
  {1998}),\ \Eprint{http://arxiv.org/abs/hep-th/9805046}{arXiv:hep-th/9805046}%
  \bibAnnoteFile{NoStop}{Konechny:1998vc}%
\bibitem{Porrati:1997ej}%
  \BibitemOpen
  \bibfield{author}{%
  \bibinfo {author} {\bibfnamefont{Massimo}\ \bibnamefont{Porrati}}\ and\
  \bibinfo {author} {\bibfnamefont{Alexander}\ \bibnamefont{Rozenberg}},\ }%
  \bibfield{title}{%
  \enquote{\bibinfo {title} {{Bound states at threshold in supersymmetric
  quantum mechanics}},}\ }%
  \bibfield{journal}{%
  \Doi{10.1016/S0550-3213(97)00804-3}{\bibinfo {journal} {Nucl. Phys. B}}\ }%
  \textbf{\bibinfo {volume} {515}},\ \bibinfo {pages} {184--202} (\bibinfo
  {year} {1998}),\
  \Eprint{http://arxiv.org/abs/hep-th/9708119}{arXiv:hep-th/9708119}%
  \bibAnnoteFile{NoStop}{Porrati:1997ej}%
\bibitem{Sethi:2000zf}%
  \BibitemOpen
  \bibfield{author}{%
  \bibinfo {author} {\bibfnamefont{Savdeep}\ \bibnamefont{Sethi}}\ and\
  \bibinfo {author} {\bibfnamefont{Mark}\ \bibnamefont{Stern}},\ }%
  \bibfield{title}{%
  \enquote{\bibinfo {title} {{Invariance theorems for supersymmetric Yang-Mills
  theories}},}\ }%
  \bibfield{journal}{%
  \Doi{10.4310/ATMP.2000.v4.n2.a8}{\bibinfo {journal} {Adv. Theor. Math.
  Phys.}}\ }%
  \textbf{\bibinfo {volume} {4}},\ \bibinfo {pages} {487--501} (\bibinfo {year}
  {2000}),\ \Eprint{http://arxiv.org/abs/hep-th/0001189}{arXiv:hep-th/0001189}%
  \bibAnnoteFile{NoStop}{Sethi:2000zf}%
\bibitem{Lin:2014wka}%
  \BibitemOpen
  \bibfield{author}{%
  \bibinfo {author} {\bibfnamefont{Ying-Hsuan}\ \bibnamefont{Lin}}\ and\
  \bibinfo {author} {\bibfnamefont{Xi}~\bibnamefont{Yin}},\ }%
  \bibfield{title}{%
  \enquote{\bibinfo {title} {{On the Ground State Wave Function of Matrix
  Theory}},}\ }%
  \bibfield{journal}{%
  \Doi{10.1007/JHEP11(2015)027}{\bibinfo {journal} {JHEP}}\ }%
  \textbf{\bibinfo {volume} {11}},\ \bibinfo {pages} {027} (\bibinfo {year}
  {2015}),\ \Eprint{http://arxiv.org/abs/1402.0055}{arXiv:1402.0055 [hep-th]}%
  \bibAnnoteFile{NoStop}{Lin:2014wka}%
\bibitem{Douglas:1996yp}%
  \BibitemOpen
  \bibfield{author}{%
  \bibinfo {author} {\bibfnamefont{Michael~R.}\ \bibnamefont{Douglas}},
  \bibinfo {author} {\bibfnamefont{Daniel~N.}\ \bibnamefont{Kabat}}, \bibinfo
  {author} {\bibfnamefont{Philippe}\ \bibnamefont{Pouliot}},\ and\ \bibinfo
  {author} {\bibfnamefont{Stephen~H.}\ \bibnamefont{Shenker}},\ }%
  \bibfield{title}{%
  \enquote{\bibinfo {title} {{D-branes and short distances in string
  theory}},}\ }%
  \bibfield{journal}{%
  \Doi{10.1016/S0550-3213(96)00619-0}{\bibinfo {journal} {Nucl. Phys. B}}\ }%
  \textbf{\bibinfo {volume} {485}},\ \bibinfo {pages} {85--127} (\bibinfo
  {year} {1997}),\
  \Eprint{http://arxiv.org/abs/hep-th/9608024}{arXiv:hep-th/9608024}%
  \bibAnnoteFile{NoStop}{Douglas:1996yp}%
\bibitem{Becker:1997wh}%
  \BibitemOpen
  \bibfield{author}{%
  \bibinfo {author} {\bibfnamefont{Katrin}\ \bibnamefont{Becker}}\ and\
  \bibinfo {author} {\bibfnamefont{Melanie}\ \bibnamefont{Becker}},\ }%
  \bibfield{title}{%
  \enquote{\bibinfo {title} {{A Two loop test of M(atrix) theory}},}\ }%
  \bibfield{journal}{%
  \Doi{10.1016/S0550-3213(97)00518-X}{\bibinfo {journal} {Nucl. Phys. B}}\ }%
  \textbf{\bibinfo {volume} {506}},\ \bibinfo {pages} {48--60} (\bibinfo {year}
  {1997}),\ \Eprint{http://arxiv.org/abs/hep-th/9705091}{arXiv:hep-th/9705091}%
  \bibAnnoteFile{NoStop}{Becker:1997wh}%
\bibitem{Becker:1997xw}%
  \BibitemOpen
  \bibfield{author}{%
  \bibinfo {author} {\bibfnamefont{Katrin}\ \bibnamefont{Becker}}, \bibinfo
  {author} {\bibfnamefont{Melanie}\ \bibnamefont{Becker}}, \bibinfo {author}
  {\bibfnamefont{Joseph}\ \bibnamefont{Polchinski}},\ and\ \bibinfo {author}
  {\bibfnamefont{Arkady~A.}\ \bibnamefont{Tseytlin}},\ }%
  \bibfield{title}{%
  \enquote{\bibinfo {title} {{Higher order graviton scattering in M(atrix)
  theory}},}\ }%
  \bibfield{journal}{%
  \Doi{10.1103/PhysRevD.56.R3174}{\bibinfo {journal} {Phys. Rev. D}}\ }%
  \textbf{\bibinfo {volume} {56}},\ \bibinfo {pages} {R3174--R3178} (\bibinfo
  {year} {1997}),\
  \Eprint{http://arxiv.org/abs/hep-th/9706072}{arXiv:hep-th/9706072}%
  \bibAnnoteFile{NoStop}{Becker:1997xw}%
\bibitem{Polchinski:1997pz}%
  \BibitemOpen
  \bibfield{author}{%
  \bibinfo {author} {\bibfnamefont{Joseph}\ \bibnamefont{Polchinski}}\ and\
  \bibinfo {author} {\bibfnamefont{Philippe}\ \bibnamefont{Pouliot}},\ }%
  \bibfield{title}{%
  \enquote{\bibinfo {title} {{Membrane scattering with M momentum transfer}},}\
  }%
  \bibfield{journal}{%
  \Doi{10.1103/PhysRevD.56.6601}{\bibinfo {journal} {Phys. Rev. D}}\ }%
  \textbf{\bibinfo {volume} {56}},\ \bibinfo {pages} {6601--6606} (\bibinfo
  {year} {1997}),\
  \Eprint{http://arxiv.org/abs/hep-th/9704029}{arXiv:hep-th/9704029}%
  \bibAnnoteFile{NoStop}{Polchinski:1997pz}%
\bibitem{Plefka:1998ed}%
  \BibitemOpen
  \bibfield{author}{%
  \bibinfo {author} {\bibfnamefont{Jan~C.}\ \bibnamefont{Plefka}}, \bibinfo
  {author} {\bibfnamefont{Marco}\ \bibnamefont{Serone}},\ and\ \bibinfo
  {author} {\bibfnamefont{Andrew~K.}\ \bibnamefont{Waldron}},\ }%
  \bibfield{title}{%
  \enquote{\bibinfo {title} {{The Matrix theory S matrix}},}\ }%
  \bibfield{journal}{%
  \Doi{10.1103/PhysRevLett.81.2866}{\bibinfo {journal} {Phys. Rev. Lett.}}\ }%
  \textbf{\bibinfo {volume} {81}},\ \bibinfo {pages} {2866--2869} (\bibinfo
  {year} {1998}),\
  \Eprint{http://arxiv.org/abs/hep-th/9806081}{arXiv:hep-th/9806081}%
  \bibAnnoteFile{NoStop}{Plefka:1998ed}%
\bibitem{Taylor:1996ik}%
  \BibitemOpen
  \bibfield{author}{%
  \bibinfo {author} {\bibfnamefont{Washington}\ \bibnamefont{Taylor}},\ }%
  \bibfield{title}{%
  \enquote{\bibinfo {title} {{D-brane field theory on compact spaces}},}\ }%
  \bibfield{journal}{%
  \Doi{10.1016/S0370-2693(97)00033-6}{\bibinfo {journal} {Phys. Lett. B}}\ }%
  \textbf{\bibinfo {volume} {394}},\ \bibinfo {pages} {283--287} (\bibinfo
  {year} {1997}),\
  \Eprint{http://arxiv.org/abs/hep-th/9611042}{arXiv:hep-th/9611042}%
  \bibAnnoteFile{NoStop}{Taylor:1996ik}%
\bibitem{Dijkgraaf:1997vv}%
  \BibitemOpen
  \bibfield{author}{%
  \bibinfo {author} {\bibfnamefont{Robbert}\ \bibnamefont{Dijkgraaf}}, \bibinfo
  {author} {\bibfnamefont{Erik~P.}\ \bibnamefont{Verlinde}},\ and\ \bibinfo
  {author} {\bibfnamefont{Herman~L.}\ \bibnamefont{Verlinde}},\ }%
  \bibfield{title}{%
  \enquote{\bibinfo {title} {{Matrix string theory}},}\ }%
  \bibfield{journal}{%
  \Doi{10.1016/S0550-3213(97)00326-X}{\bibinfo {journal} {Nucl. Phys. B}}\ }%
  \textbf{\bibinfo {volume} {500}},\ \bibinfo {pages} {43--61} (\bibinfo {year}
  {1997}),\ \Eprint{http://arxiv.org/abs/hep-th/9703030}{arXiv:hep-th/9703030}%
  \bibAnnoteFile{NoStop}{Dijkgraaf:1997vv}%
\bibitem{Bergman:1997yw}%
  \BibitemOpen
  \bibfield{author}{%
  \bibinfo {author} {\bibfnamefont{Oren}\ \bibnamefont{Bergman}},\ }%
  \bibfield{title}{%
  \enquote{\bibinfo {title} {{Three pronged strings and 1/4 BPS states in N=4
  superYang-Mills theory}},}\ }%
  \bibfield{journal}{%
  \Doi{10.1016/S0550-3213(98)00345-9}{\bibinfo {journal} {Nucl. Phys. B}}\ }%
  \textbf{\bibinfo {volume} {525}},\ \bibinfo {pages} {104--116} (\bibinfo
  {year} {1998}),\
  \Eprint{http://arxiv.org/abs/hep-th/9712211}{arXiv:hep-th/9712211}%
  \bibAnnoteFile{NoStop}{Bergman:1997yw}%
\bibitem{Bergman:1998gs}%
  \BibitemOpen
  \bibfield{author}{%
  \bibinfo {author} {\bibfnamefont{Oren}\ \bibnamefont{Bergman}}\ and\ \bibinfo
  {author} {\bibfnamefont{Barak}\ \bibnamefont{Kol}},\ }%
  \bibfield{title}{%
  \enquote{\bibinfo {title} {{String webs and 1/4 BPS monopoles}},}\ }%
  \bibfield{journal}{%
  \Doi{10.1016/S0550-3213(98)00565-3}{\bibinfo {journal} {Nucl. Phys. B}}\ }%
  \textbf{\bibinfo {volume} {536}},\ \bibinfo {pages} {149--174} (\bibinfo
  {year} {1998}),\
  \Eprint{http://arxiv.org/abs/hep-th/9804160}{arXiv:hep-th/9804160}%
  \bibAnnoteFile{NoStop}{Bergman:1998gs}%
\bibitem{Schwarz:1996bh}%
  \BibitemOpen
  \bibfield{author}{%
  \bibinfo {author} {\bibfnamefont{John~H.}\ \bibnamefont{Schwarz}},\ }%
  \bibfield{title}{%
  \enquote{\bibinfo {title} {{Lectures on superstring and M theory dualities:
  Given at ICTP Spring School and at TASI Summer School}},}\ }%
  \bibfield{journal}{%
  \Doi{10.1016/S0920-5632(97)00070-4}{\bibinfo {journal} {Nucl. Phys. B Proc.
  Suppl.}}\ }%
  \textbf{\bibinfo {volume} {55}},\ \bibinfo {pages} {1--32} (\bibinfo {year}
  {1997}),\ \Eprint{http://arxiv.org/abs/hep-th/9607201}{arXiv:hep-th/9607201}%
  \bibAnnoteFile{NoStop}{Schwarz:1996bh}%
\bibitem{Aharony:1997ju}%
  \BibitemOpen
  \bibfield{author}{%
  \bibinfo {author} {\bibfnamefont{Ofer}\ \bibnamefont{Aharony}}\ and\ \bibinfo
  {author} {\bibfnamefont{Amihay}\ \bibnamefont{Hanany}},\ }%
  \bibfield{title}{%
  \enquote{\bibinfo {title} {{Branes, superpotentials and superconformal fixed
  points}},}\ }%
  \bibfield{journal}{%
  \Doi{10.1016/S0550-3213(97)00472-0}{\bibinfo {journal} {Nucl. Phys. B}}\ }%
  \textbf{\bibinfo {volume} {504}},\ \bibinfo {pages} {239--271} (\bibinfo
  {year} {1997}),\
  \Eprint{http://arxiv.org/abs/hep-th/9704170}{arXiv:hep-th/9704170}%
  \bibAnnoteFile{NoStop}{Aharony:1997ju}%
\bibitem{Aharony:1997bh}%
  \BibitemOpen
  \bibfield{author}{%
  \bibinfo {author} {\bibfnamefont{Ofer}\ \bibnamefont{Aharony}}, \bibinfo
  {author} {\bibfnamefont{Amihay}\ \bibnamefont{Hanany}},\ and\ \bibinfo
  {author} {\bibfnamefont{Barak}\ \bibnamefont{Kol}},\ }%
  \bibfield{title}{%
  \enquote{\bibinfo {title} {{Webs of (p,q) five-branes, five-dimensional field
  theories and grid diagrams}},}\ }%
  \bibfield{journal}{%
  \Doi{10.1088/1126-6708/1998/01/002}{\bibinfo {journal} {JHEP}}\ }%
  \textbf{\bibinfo {volume} {01}},\ \bibinfo {pages} {002} (\bibinfo {year}
  {1998}),\ \Eprint{http://arxiv.org/abs/hep-th/9710116}{arXiv:hep-th/9710116}%
  \bibAnnoteFile{NoStop}{Aharony:1997bh}%
\bibitem{Dasgupta:1997pu}%
  \BibitemOpen
  \bibfield{author}{%
  \bibinfo {author} {\bibfnamefont{Keshav}\ \bibnamefont{Dasgupta}}\ and\
  \bibinfo {author} {\bibfnamefont{Sunil}\ \bibnamefont{Mukhi}},\ }%
  \bibfield{title}{%
  \enquote{\bibinfo {title} {{BPS nature of three string junctions}},}\ }%
  \bibfield{journal}{%
  \Doi{10.1016/S0370-2693(98)00140-3}{\bibinfo {journal} {Phys. Lett. B}}\ }%
  \textbf{\bibinfo {volume} {423}},\ \bibinfo {pages} {261--264} (\bibinfo
  {year} {1998}),\
  \Eprint{http://arxiv.org/abs/hep-th/9711094}{arXiv:hep-th/9711094}%
  \bibAnnoteFile{NoStop}{Dasgupta:1997pu}%
\bibitem{Sen:1997xi}%
  \BibitemOpen
  \bibfield{author}{%
  \bibinfo {author} {\bibfnamefont{Ashoke}\ \bibnamefont{Sen}},\ }%
  \bibfield{title}{%
  \enquote{\bibinfo {title} {{String network}},}\ }%
  \bibfield{journal}{%
  \Doi{10.1088/1126-6708/1998/03/005}{\bibinfo {journal} {JHEP}}\ }%
  \textbf{\bibinfo {volume} {03}},\ \bibinfo {pages} {005} (\bibinfo {year}
  {1998}),\ \Eprint{http://arxiv.org/abs/hep-th/9711130}{arXiv:hep-th/9711130}%
  \bibAnnoteFile{NoStop}{Sen:1997xi}%
\bibitem{Kol:1998zb}%
  \BibitemOpen
  \bibfield{author}{%
  \bibinfo {author} {\bibfnamefont{Barak}\ \bibnamefont{Kol}},\ }%
  \bibfield{title}{%
  \enquote{\bibinfo {title} {{Thermal monopoles}},}\ }%
  \bibfield{journal}{%
  \Doi{10.1088/1126-6708/2000/07/026}{\bibinfo {journal} {JHEP}}\ }%
  \textbf{\bibinfo {volume} {07}},\ \bibinfo {pages} {026} (\bibinfo {year}
  {2000}),\ \Eprint{http://arxiv.org/abs/hep-th/9812021}{arXiv:hep-th/9812021}%
  \bibAnnoteFile{NoStop}{Kol:1998zb}%
\bibitem{Banerjee:2008pu}%
  \BibitemOpen
  \bibfield{author}{%
  \bibinfo {author} {\bibfnamefont{Shamik}\ \bibnamefont{Banerjee}}, \bibinfo
  {author} {\bibfnamefont{Ashoke}\ \bibnamefont{Sen}},\ and\ \bibinfo {author}
  {\bibfnamefont{Yogesh~K.}\ \bibnamefont{Srivastava}},\ }%
  \bibfield{title}{%
  \enquote{\bibinfo {title} {{Partition Functions of Torsion \ensuremath{>} 1
  Dyons in Heterotic String Theory on T**6}},}\ }%
  \bibfield{journal}{%
  \Doi{10.1088/1126-6708/2008/05/098}{\bibinfo {journal} {JHEP}}\ }%
  \textbf{\bibinfo {volume} {05}},\ \bibinfo {pages} {098} (\bibinfo {year}
  {2008}),\ \Eprint{http://arxiv.org/abs/0802.1556}{arXiv:0802.1556 [hep-th]}%
  \bibAnnoteFile{NoStop}{Banerjee:2008pu}%
\bibitem{Sen:2008ht}%
  \BibitemOpen
  \bibfield{author}{%
  \bibinfo {author} {\bibfnamefont{Ashoke}\ \bibnamefont{Sen}},\ }%
  \bibfield{title}{%
  \enquote{\bibinfo {title} {{Wall Crossing Formula for N=4 Dyons: A
  Macroscopic Derivation}},}\ }%
  \bibfield{journal}{%
  \Doi{10.1088/1126-6708/2008/07/078}{\bibinfo {journal} {JHEP}}\ }%
  \textbf{\bibinfo {volume} {07}},\ \bibinfo {pages} {078} (\bibinfo {year}
  {2008}),\ \Eprint{http://arxiv.org/abs/0803.3857}{arXiv:0803.3857 [hep-th]}%
  \bibAnnoteFile{NoStop}{Sen:2008ht}%
\bibitem{Sen:2012hv}%
  \BibitemOpen
  \bibfield{author}{%
  \bibinfo {author} {\bibfnamefont{Ashoke}\ \bibnamefont{Sen}},\ }%
  \bibfield{title}{%
  \enquote{\bibinfo {title} {{BPS Spectrum, Indices and Wall Crossing in N=4
  Supersymmetric Yang-Mills Theories}},}\ }%
  \bibfield{journal}{%
  \Doi{10.1007/JHEP06(2012)164}{\bibinfo {journal} {JHEP}}\ }%
  \textbf{\bibinfo {volume} {06}},\ \bibinfo {pages} {164} (\bibinfo {year}
  {2012}),\ \Eprint{http://arxiv.org/abs/1203.4889}{arXiv:1203.4889 [hep-th]}%
  \bibAnnoteFile{NoStop}{Sen:2012hv}%
\bibitem{Motl:1997th}%
  \BibitemOpen
  \bibfield{author}{%
  \bibinfo {author} {\bibfnamefont{Lubos}\ \bibnamefont{Motl}},\ }%
  \bibfield{title}{%
  \enquote{\bibinfo {title} {{Proposals on nonperturbative superstring
  interactions}},}\ }%
   (\bibinfo {month} {1}\ \bibinfo {year} {1997}),\
  \Eprint{http://arxiv.org/abs/hep-th/9701025}{arXiv:hep-th/9701025}%
  \bibAnnoteFile{NoStop}{Motl:1997th}%
\bibitem{Giddings:1998yd}%
  \BibitemOpen
  \bibfield{author}{%
  \bibinfo {author} {\bibfnamefont{Steven~B.}\ \bibnamefont{Giddings}},
  \bibinfo {author} {\bibfnamefont{Feike}\ \bibnamefont{Hacquebord}},\ and\
  \bibinfo {author} {\bibfnamefont{Herman~L.}\ \bibnamefont{Verlinde}},\ }%
  \bibfield{title}{%
  \enquote{\bibinfo {title} {{High-energy scattering and D pair creation in
  matrix string theory}},}\ }%
  \bibfield{journal}{%
  \Doi{10.1016/S0550-3213(98)00662-2}{\bibinfo {journal} {Nucl. Phys. B}}\ }%
  \textbf{\bibinfo {volume} {537}},\ \bibinfo {pages} {260--296} (\bibinfo
  {year} {1999}),\
  \Eprint{http://arxiv.org/abs/hep-th/9804121}{arXiv:hep-th/9804121}%
  \bibAnnoteFile{NoStop}{Giddings:1998yd}%
\bibitem{Bonelli:1998wx}%
  \BibitemOpen
  \bibfield{author}{%
  \bibinfo {author} {\bibfnamefont{G.}~\bibnamefont{Bonelli}}, \bibinfo
  {author} {\bibfnamefont{L.}~\bibnamefont{Bonora}},\ and\ \bibinfo {author}
  {\bibfnamefont{F.}~\bibnamefont{Nesti}},\ }%
  \bibfield{title}{%
  \enquote{\bibinfo {title} {{String interactions from matrix string
  theory}},}\ }%
  \bibfield{journal}{%
  \Doi{10.1016/S0550-3213(98)00729-9}{\bibinfo {journal} {Nucl. Phys. B}}\ }%
  \textbf{\bibinfo {volume} {538}},\ \bibinfo {pages} {100--116} (\bibinfo
  {year} {1999}),\
  \Eprint{http://arxiv.org/abs/hep-th/9807232}{arXiv:hep-th/9807232}%
  \bibAnnoteFile{NoStop}{Bonelli:1998wx}%
\bibitem{Bonelli:1999qa}%
  \BibitemOpen
  \bibfield{author}{%
  \bibinfo {author} {\bibfnamefont{G.}~\bibnamefont{Bonelli}}, \bibinfo
  {author} {\bibfnamefont{L.}~\bibnamefont{Bonora}}, \bibinfo {author}
  {\bibfnamefont{F.}~\bibnamefont{Nesti}},\ and\ \bibinfo {author}
  {\bibfnamefont{Alessandro}\ \bibnamefont{Tomasiello}},\ }%
  \bibfield{title}{%
  \enquote{\bibinfo {title} {{Matrix string theory and its moduli space}},}\ }%
  \bibfield{journal}{%
  \Doi{10.1016/S0550-3213(99)00271-0}{\bibinfo {journal} {Nucl. Phys. B}}\ }%
  \textbf{\bibinfo {volume} {554}},\ \bibinfo {pages} {103--135} (\bibinfo
  {year} {1999}),\
  \Eprint{http://arxiv.org/abs/hep-th/9901093}{arXiv:hep-th/9901093}%
  \bibAnnoteFile{NoStop}{Bonelli:1999qa}%
\bibitem{Grisaru:1977px}%
  \BibitemOpen
  \bibfield{author}{%
  \bibinfo {author} {\bibfnamefont{Marcus~T.}\ \bibnamefont{Grisaru}}\ and\
  \bibinfo {author} {\bibfnamefont{H.~N.}\ \bibnamefont{Pendleton}},\ }%
  \bibfield{title}{%
  \enquote{\bibinfo {title} {{Some Properties of Scattering Amplitudes in
  Supersymmetric Theories}},}\ }%
  \bibfield{journal}{%
  \Doi{10.1016/0550-3213(77)90277-2}{\bibinfo {journal} {Nucl. Phys. B}}\ }%
  \textbf{\bibinfo {volume} {124}},\ \bibinfo {pages} {81--92} (\bibinfo {year}
  {1977})%
  \bibAnnoteFile{NoStop}{Grisaru:1977px}%
\bibitem{Nair:1988bq}%
  \BibitemOpen
  \bibfield{author}{%
  \bibinfo {author} {\bibfnamefont{V.~P.}\ \bibnamefont{Nair}},\ }%
  \bibfield{title}{%
  \enquote{\bibinfo {title} {{A Current Algebra for Some Gauge Theory
  Amplitudes}},}\ }%
  \bibfield{journal}{%
  \Doi{10.1016/0370-2693(88)91471-2}{\bibinfo {journal} {Phys. Lett. B}}\ }%
  \textbf{\bibinfo {volume} {214}},\ \bibinfo {pages} {215--218} (\bibinfo
  {year} {1988})%
  \bibAnnoteFile{NoStop}{Nair:1988bq}%
\bibitem{Witten:2003nn}%
  \BibitemOpen
  \bibfield{author}{%
  \bibinfo {author} {\bibfnamefont{Edward}\ \bibnamefont{Witten}},\ }%
  \bibfield{title}{%
  \enquote{\bibinfo {title} {{Perturbative gauge theory as a string theory in
  twistor space}},}\ }%
  \bibfield{journal}{%
  \Doi{10.1007/s00220-004-1187-3}{\bibinfo {journal} {Commun. Math. Phys.}}\ }%
  \textbf{\bibinfo {volume} {252}},\ \bibinfo {pages} {189--258} (\bibinfo
  {year} {2004}),\
  \Eprint{http://arxiv.org/abs/hep-th/0312171}{arXiv:hep-th/0312171}%
  \bibAnnoteFile{NoStop}{Witten:2003nn}%
\bibitem{Bianchi:2008pu}%
  \BibitemOpen
  \bibfield{author}{%
  \bibinfo {author} {\bibfnamefont{Massimo}\ \bibnamefont{Bianchi}}, \bibinfo
  {author} {\bibfnamefont{Henriette}\ \bibnamefont{Elvang}},\ and\ \bibinfo
  {author} {\bibfnamefont{Daniel~Z.}\ \bibnamefont{Freedman}},\ }%
  \bibfield{title}{%
  \enquote{\bibinfo {title} {{Generating Tree Amplitudes in N=4 SYM and N = 8
  SG}},}\ }%
  \bibfield{journal}{%
  \Doi{10.1088/1126-6708/2008/09/063}{\bibinfo {journal} {JHEP}}\ }%
  \textbf{\bibinfo {volume} {09}},\ \bibinfo {pages} {063} (\bibinfo {year}
  {2008}),\ \Eprint{http://arxiv.org/abs/0805.0757}{arXiv:0805.0757 [hep-th]}%
  \bibAnnoteFile{NoStop}{Bianchi:2008pu}%
\bibitem{Arkani-Hamed:2008owk}%
  \BibitemOpen
  \bibfield{author}{%
  \bibinfo {author} {\bibfnamefont{Nima}\ \bibnamefont{Arkani-Hamed}}, \bibinfo
  {author} {\bibfnamefont{Freddy}\ \bibnamefont{Cachazo}},\ and\ \bibinfo
  {author} {\bibfnamefont{Jared}\ \bibnamefont{Kaplan}},\ }%
  \bibfield{title}{%
  \enquote{\bibinfo {title} {{What is the Simplest Quantum Field Theory?}}.}\
  }%
  \bibfield{journal}{%
  \Doi{10.1007/JHEP09(2010)016}{\bibinfo {journal} {JHEP}}\ }%
  \textbf{\bibinfo {volume} {09}},\ \bibinfo {pages} {016} (\bibinfo {year}
  {2010}),\ \Eprint{http://arxiv.org/abs/0808.1446}{arXiv:0808.1446 [hep-th]}%
  \bibAnnoteFile{NoStop}{Arkani-Hamed:2008owk}%
\bibitem{Arkani-Hamed:2008bsc}%
  \BibitemOpen
  \bibfield{author}{%
  \bibinfo {author} {\bibfnamefont{Nima}\ \bibnamefont{Arkani-Hamed}}\ and\
  \bibinfo {author} {\bibfnamefont{Jared}\ \bibnamefont{Kaplan}},\ }%
  \bibfield{title}{%
  \enquote{\bibinfo {title} {{On Tree Amplitudes in Gauge Theory and
  Gravity}},}\ }%
  \bibfield{journal}{%
  \Doi{10.1088/1126-6708/2008/04/076}{\bibinfo {journal} {JHEP}}\ }%
  \textbf{\bibinfo {volume} {04}},\ \bibinfo {pages} {076} (\bibinfo {year}
  {2008}),\ \Eprint{http://arxiv.org/abs/0801.2385}{arXiv:0801.2385 [hep-th]}%
  \bibAnnoteFile{NoStop}{Arkani-Hamed:2008bsc}%
\bibitem{Huang:2011um}%
  \BibitemOpen
  \bibfield{author}{%
  \bibinfo {author} {\bibfnamefont{Yu-tin}\ \bibnamefont{Huang}},\ }%
  \bibfield{title}{%
  \enquote{\bibinfo {title} {{Non-Chiral S-Matrix of N=4 Super Yang-Mills}},}\
  }%
   (\bibinfo {month} {4}\ \bibinfo {year} {2011}),\
  \Eprint{http://arxiv.org/abs/1104.2021}{arXiv:1104.2021 [hep-th]}%
  \bibAnnoteFile{NoStop}{Huang:2011um}%
\bibitem{Taylor:1997dy}%
  \BibitemOpen
  \bibfield{author}{%
  \bibinfo {author} {\bibfnamefont{Washington}\ \bibnamefont{Taylor}},\ }%
  \enquote{\bibinfo {title} {{Lectures on D-branes, gauge theory and
  M(atrices)}},}\ in\ \emph{\bibinfo {booktitle} {{2nd Trieste Conference on
  Duality in String Theory}}}\ (\bibinfo {year} {1997})\ pp.\ \bibinfo {pages}
  {192--271},\
  \Eprint{http://arxiv.org/abs/hep-th/9801182}{arXiv:hep-th/9801182}%
  \bibAnnoteFile{NoStop}{Taylor:1997dy}%
\bibitem{Witten:1995im}%
  \BibitemOpen
  \bibfield{author}{%
  \bibinfo {author} {\bibfnamefont{Edward}\ \bibnamefont{Witten}},\ }%
  \bibfield{title}{%
  \enquote{\bibinfo {title} {{Bound states of strings and p-branes}},}\ }%
  \bibfield{journal}{%
  \Doi{10.1016/0550-3213(95)00610-9}{\bibinfo {journal} {Nucl. Phys. B}}\ }%
  \textbf{\bibinfo {volume} {460}},\ \bibinfo {pages} {335--350} (\bibinfo
  {year} {1996}),\
  \Eprint{http://arxiv.org/abs/hep-th/9510135}{arXiv:hep-th/9510135}%
  \bibAnnoteFile{NoStop}{Witten:1995im}%
\bibitem{Plefka:1997xq}%
  \BibitemOpen
  \bibfield{author}{%
  \bibinfo {author} {\bibfnamefont{Jan}\ \bibnamefont{Plefka}}\ and\ \bibinfo
  {author} {\bibfnamefont{Andrew}\ \bibnamefont{Waldron}},\ }%
  \bibfield{title}{%
  \enquote{\bibinfo {title} {{On the quantum mechanics of M(atrix) theory}},}\
  }%
  \bibfield{journal}{%
  \Doi{10.1016/S0550-3213(97)00762-1}{\bibinfo {journal} {Nucl. Phys. B}}\ }%
  \textbf{\bibinfo {volume} {512}},\ \bibinfo {pages} {460--484} (\bibinfo
  {year} {1998}),\
  \Eprint{http://arxiv.org/abs/hep-th/9710104}{arXiv:hep-th/9710104}%
  \bibAnnoteFile{NoStop}{Plefka:1997xq}%
\bibitem{Nahm:1979yw}%
  \BibitemOpen
  \bibfield{author}{%
  \bibinfo {author} {\bibfnamefont{W.}~\bibnamefont{Nahm}},\ }%
  \bibfield{title}{%
  \enquote{\bibinfo {title} {{A Simple Formalism for the BPS Monopole}},}\ }%
  \bibfield{journal}{%
  \Doi{10.1016/0370-2693(80)90961-2}{\bibinfo {journal} {Phys. Lett. B}}\ }%
  \textbf{\bibinfo {volume} {90}},\ \bibinfo {pages} {413--414} (\bibinfo
  {year} {1980})%
  \bibAnnoteFile{NoStop}{Nahm:1979yw}%
\bibitem{Diaconescu:1996rk}%
  \BibitemOpen
  \bibfield{author}{%
  \bibinfo {author} {\bibfnamefont{Duiliu-Emanuel}\ \bibnamefont{Diaconescu}},\
  }%
  \bibfield{title}{%
  \enquote{\bibinfo {title} {{D-branes, monopoles and Nahm equations}},}\ }%
  \bibfield{journal}{%
  \Doi{10.1016/S0550-3213(97)00438-0}{\bibinfo {journal} {Nucl. Phys. B}}\ }%
  \textbf{\bibinfo {volume} {503}},\ \bibinfo {pages} {220--238} (\bibinfo
  {year} {1997}),\
  \Eprint{http://arxiv.org/abs/hep-th/9608163}{arXiv:hep-th/9608163}%
  \bibAnnoteFile{NoStop}{Diaconescu:1996rk}%
\bibitem{Antoniadis:1993ze}%
  \BibitemOpen
  \bibfield{author}{%
  \bibinfo {author} {\bibfnamefont{Ignatios}\ \bibnamefont{Antoniadis}},
  \bibinfo {author} {\bibfnamefont{E.}~\bibnamefont{Gava}}, \bibinfo {author}
  {\bibfnamefont{K.~S.}\ \bibnamefont{Narain}},\ and\ \bibinfo {author}
  {\bibfnamefont{T.~R.}\ \bibnamefont{Taylor}},\ }%
  \bibfield{title}{%
  \enquote{\bibinfo {title} {{Topological amplitudes in string theory}},}\ }%
  \bibfield{journal}{%
  \Doi{10.1016/0550-3213(94)90617-3}{\bibinfo {journal} {Nucl. Phys. B}}\ }%
  \textbf{\bibinfo {volume} {413}},\ \bibinfo {pages} {162--184} (\bibinfo
  {year} {1994}),\
  \Eprint{http://arxiv.org/abs/hep-th/9307158}{arXiv:hep-th/9307158}%
  \bibAnnoteFile{NoStop}{Antoniadis:1993ze}%
\bibitem{Alday:2007he}%
  \BibitemOpen
  \bibfield{author}{%
  \bibinfo {author} {\bibfnamefont{Luis~F.}\ \bibnamefont{Alday}}\ and\
  \bibinfo {author} {\bibfnamefont{Juan}\ \bibnamefont{Maldacena}},\ }%
  \bibfield{title}{%
  \enquote{\bibinfo {title} {{Comments on gluon scattering amplitudes via
  AdS/CFT}},}\ }%
  \bibfield{journal}{%
  \Doi{10.1088/1126-6708/2007/11/068}{\bibinfo {journal} {JHEP}}\ }%
  \textbf{\bibinfo {volume} {11}},\ \bibinfo {pages} {068} (\bibinfo {year}
  {2007}),\ \Eprint{http://arxiv.org/abs/0710.1060}{arXiv:0710.1060 [hep-th]}%
  \bibAnnoteFile{NoStop}{Alday:2007he}%
\bibitem{Weinberg:1965nx}%
  \BibitemOpen
  \bibfield{author}{%
  \bibinfo {author} {\bibfnamefont{Steven}\ \bibnamefont{Weinberg}},\ }%
  \bibfield{title}{%
  \enquote{\bibinfo {title} {{Infrared photons and gravitons}},}\ }%
  \bibfield{journal}{%
  \Doi{10.1103/PhysRev.140.B516}{\bibinfo {journal} {Phys. Rev.}}\ }%
  \textbf{\bibinfo {volume} {140}},\ \bibinfo {pages} {B516--B524} (\bibinfo
  {year} {1965})%
  \bibAnnoteFile{NoStop}{Weinberg:1965nx}%
\bibitem{Cachazo:2014fwa}%
  \BibitemOpen
  \bibfield{author}{%
  \bibinfo {author} {\bibfnamefont{Freddy}\ \bibnamefont{Cachazo}}\ and\
  \bibinfo {author} {\bibfnamefont{Andrew}\ \bibnamefont{Strominger}},\ }%
  \bibfield{title}{%
  \enquote{\bibinfo {title} {{Evidence for a New Soft Graviton Theorem}},}\ }%
   (\bibinfo {month} {4}\ \bibinfo {year} {2014}),\
  \Eprint{http://arxiv.org/abs/1404.4091}{arXiv:1404.4091 [hep-th]}%
  \bibAnnoteFile{NoStop}{Cachazo:2014fwa}%
\bibitem{OurSoft}%
  \BibitemOpen
  \bibfield{author}{%
  \bibinfo {author} {\bibfnamefont{Aidan}\ \bibnamefont{Herderschee}}\ and\
  \bibinfo {author} {\bibfnamefont{Juan}\ \bibnamefont{Maldacena}},\ }%
  \enquote{\bibinfo {title} {Soft theorems in matrix theory},}\ \bibinfo
  {howpublished} {to appear}%
  \bibAnnoteFile{NoStop}{OurSoft}%
\bibitem{Banks:1997it}%
  \BibitemOpen
  \bibfield{author}{%
  \bibinfo {author} {\bibfnamefont{Tom}\ \bibnamefont{Banks}}\ and\ \bibinfo
  {author} {\bibfnamefont{Lubos}\ \bibnamefont{Motl}},\ }%
  \bibfield{title}{%
  \enquote{\bibinfo {title} {{Heterotic strings from matrices}},}\ }%
  \bibfield{journal}{%
  \Doi{10.1088/1126-6708/1997/12/004}{\bibinfo {journal} {JHEP}}\ }%
  \textbf{\bibinfo {volume} {12}},\ \bibinfo {pages} {004} (\bibinfo {year}
  {1997}),\ \Eprint{http://arxiv.org/abs/hep-th/9703218}{arXiv:hep-th/9703218}%
  \bibAnnoteFile{NoStop}{Banks:1997it}%
\bibitem{Motl:1997tb}%
  \BibitemOpen
  \bibfield{author}{%
  \bibinfo {author} {\bibfnamefont{Lubos}\ \bibnamefont{Motl}}\ and\ \bibinfo
  {author} {\bibfnamefont{Leonard}\ \bibnamefont{Susskind}},\ }%
  \bibfield{title}{%
  \enquote{\bibinfo {title} {{Finite N heterotic matrix models and discrete
  light cone quantization}},}\ }%
   (\bibinfo {month} {8}\ \bibinfo {year} {1997}),\
  \Eprint{http://arxiv.org/abs/hep-th/9708083}{arXiv:hep-th/9708083}%
  \bibAnnoteFile{NoStop}{Motl:1997tb}%
\bibitem{Britto:2005fq}%
  \BibitemOpen
  \bibfield{author}{%
  \bibinfo {author} {\bibfnamefont{Ruth}\ \bibnamefont{Britto}}, \bibinfo
  {author} {\bibfnamefont{Freddy}\ \bibnamefont{Cachazo}}, \bibinfo {author}
  {\bibfnamefont{Bo}~\bibnamefont{Feng}},\ and\ \bibinfo {author}
  {\bibfnamefont{Edward}\ \bibnamefont{Witten}},\ }%
  \bibfield{title}{%
  \enquote{\bibinfo {title} {{Direct proof of tree-level recursion relation in
  Yang-Mills theory}},}\ }%
  \bibfield{journal}{%
  \Doi{10.1103/PhysRevLett.94.181602}{\bibinfo {journal} {Phys. Rev. Lett.}}\
  }%
  \textbf{\bibinfo {volume} {94}},\ \bibinfo {pages} {181602} (\bibinfo {year}
  {2005}),\ \Eprint{http://arxiv.org/abs/hep-th/0501052}{arXiv:hep-th/0501052}%
  \bibAnnoteFile{NoStop}{Britto:2005fq}%
\bibitem{Asano:2014eca}%
  \BibitemOpen
  \bibfield{author}{%
  \bibinfo {author} {\bibfnamefont{Yuhma}\ \bibnamefont{Asano}}, \bibinfo
  {author} {\bibfnamefont{Goro}\ \bibnamefont{Ishiki}},\ and\ \bibinfo {author}
  {\bibfnamefont{Shinji}\ \bibnamefont{Shimasaki}},\ }%
  \bibfield{title}{%
  \enquote{\bibinfo {title} {{Emergent bubbling geometries in gauge theories
  with SU(2|4) symmetry}},}\ }%
  \bibfield{journal}{%
  \Doi{10.1007/JHEP09(2014)137}{\bibinfo {journal} {JHEP}}\ }%
  \textbf{\bibinfo {volume} {09}},\ \bibinfo {pages} {137} (\bibinfo {year}
  {2014}),\ \Eprint{http://arxiv.org/abs/1406.1337}{arXiv:1406.1337 [hep-th]}%
  \bibAnnoteFile{NoStop}{Asano:2014eca}%
\bibitem{Asano:2012zt}%
  \BibitemOpen
  \bibfield{author}{%
  \bibinfo {author} {\bibfnamefont{Yuhma}\ \bibnamefont{Asano}}, \bibinfo
  {author} {\bibfnamefont{Goro}\ \bibnamefont{Ishiki}}, \bibinfo {author}
  {\bibfnamefont{Takashi}\ \bibnamefont{Okada}},\ and\ \bibinfo {author}
  {\bibfnamefont{Shinji}\ \bibnamefont{Shimasaki}},\ }%
  \bibfield{title}{%
  \enquote{\bibinfo {title} {{Exact results for perturbative partition
  functions of theories with SU(2|4) symmetry}},}\ }%
  \bibfield{journal}{%
  \Doi{10.1007/JHEP02(2013)148}{\bibinfo {journal} {JHEP}}\ }%
  \textbf{\bibinfo {volume} {02}},\ \bibinfo {pages} {148} (\bibinfo {year}
  {2013}),\ \Eprint{http://arxiv.org/abs/1211.0364}{arXiv:1211.0364 [hep-th]}%
  \bibAnnoteFile{NoStop}{Asano:2012zt}%
\bibitem{Berenstein:2002jq}%
  \BibitemOpen
  \bibfield{author}{%
  \bibinfo {author} {\bibfnamefont{David~Eliecer}\ \bibnamefont{Berenstein}},
  \bibinfo {author} {\bibfnamefont{Juan~Martin}\ \bibnamefont{Maldacena}},\
  and\ \bibinfo {author} {\bibfnamefont{Horatiu~Stefan}\
  \bibnamefont{Nastase}},\ }%
  \bibfield{title}{%
  \enquote{\bibinfo {title} {{Strings in flat space and pp waves from N=4
  superYang-Mills}},}\ }%
  \bibfield{journal}{%
  \Doi{10.1088/1126-6708/2002/04/013}{\bibinfo {journal} {JHEP}}\ }%
  \textbf{\bibinfo {volume} {04}},\ \bibinfo {pages} {013} (\bibinfo {year}
  {2002}),\ \Eprint{http://arxiv.org/abs/hep-th/0202021}{arXiv:hep-th/0202021}%
  \bibAnnoteFile{NoStop}{Berenstein:2002jq}%
\end{thebibliography}%


\begin{thebibliography}{10}%
\makeatletter
\providecommand \@ifxundefined [1]{%
 \ifx #1\undefined \expandafter \@firstoftwo
 \else \expandafter \@secondoftwo
\fi
}%
\providecommand \@ifnum [1]{%
 \ifnum #1\expandafter \@firstoftwo
 \else \expandafter \@secondoftwo
\fi
}%
\providecommand \enquote [1]{``#1''}%
\providecommand \bibnamefont  [1]{#1}%
\providecommand \bibfnamefont [1]{#1}%
\providecommand \citenamefont [1]{#1}%
\providecommand\href[0]{\@sanitize\@href}%
\providecommand\@href[1]{\endgroup\@@startlink{#1}\endgroup\@@href}%
\providecommand\@@href[1]{#1\@@endlink}%
\providecommand \@sanitize [0]{\begingroup\catcode`\&12\catcode`\#12\relax}%
\@ifxundefined \pdfoutput {\@firstoftwo}{%
 \@ifnum{\z@=\pdfoutput}{\@firstoftwo}{\@secondoftwo}%
}{%
 \providecommand\@@startlink[1]{\leavevmode\special{html:<a href="#1">}}%
 \providecommand\@@endlink[0]{\special{html:</a>}}%
}{%
 \providecommand\@@startlink[1]{%
  \leavevmode
  \pdfstartlink
   attr{/Border[0 0 1 ]/H/I/C[0 1 1]}%
   user{/Subtype/Link/A<</Type/Action/S/URI/URI(#1)>>}%
  \relax
 }%
 \providecommand\@@endlink[0]{\pdfendlink}%
}%
\providecommand \url  [0]{\begingroup\@sanitize \@url }%
\providecommand \@url [1]{\endgroup\@href {#1}{\urlprefix}}%
\providecommand \urlprefix [0]{URL }%
\providecommand \Eprint[0]{\href }%
\@ifxundefined \urlstyle {%
  \providecommand \doi [1]{doi:\discretionary{}{}{}#1}%
}{%
  \providecommand \doi [0]{doi:\discretionary{}{}{}\begingroup \urlstyle{rm}\Url }%
}%
\providecommand \doibase [0]{http://dx.doi.org/}%
\providecommand \Doi[1]{\href{\doibase#1}}%
\providecommand \bibAnnote [3]{%
  \BibitemShut{#1}%
  \begin{quotation}\noindent
    \textsc{Key:}\ #2\\\textsc{Annotation:}\ #3%
  \end{quotation}%
}%
\providecommand \bibAnnoteFile [2]{%
  \IfFileExists{#2}{\bibAnnote {#1} {#2} {\input{#2}}}{}%
}%
\providecommand \typeout [0]{\immediate \write \m@ne }%
\providecommand \selectlanguage [0]{\@gobble}%
\providecommand \bibinfo [0]{\@secondoftwo}%
\providecommand \bibfield [0]{\@secondoftwo}%
\providecommand \translation [1]{[#1]}%
\providecommand \BibitemOpen[0]{}%
\providecommand \bibitemStop [0]{}%
\providecommand \bibitemNoStop [0]{.\EOS\space}%
\providecommand \EOS [0]{\spacefactor3000\relax}%
\providecommand \BibitemShut [1]{\csname bibitem#1\endcsname}%
\bibitem{Banks:1996vh}%
  \BibitemOpen
  \bibfield{author}{%
  \bibinfo {author} {\bibfnamefont{Tom}\ \bibnamefont{Banks}}, \bibinfo {author} {\bibfnamefont{W.}~\bibnamefont{Fischler}}, \bibinfo {author} {\bibfnamefont{S.~H.}\ \bibnamefont{Shenker}},\ and\ \bibinfo {author} {\bibfnamefont{Leonard}\ \bibnamefont{Susskind}},\ }%
  \bibfield{title}{%
  \enquote{\bibinfo {title} {{M theory as a matrix model: A Conjecture}},}\ }%
  \bibfield{journal}{%
  \Doi{10.1103/PhysRevD.55.5112}{\bibinfo {journal} {Phys. Rev. D}}\ }%
  \textbf{\bibinfo {volume} {55}},\ \bibinfo {pages} {5112--5128} (\bibinfo {year} {1997}),\ \Eprint{http://arxiv.org/abs/hep-th/9610043}{arXiv:hep-th/9610043}%
  \bibAnnoteFile{NoStop}{Banks:1996vh}%
\bibitem{Weinberg:1965nx}%
  \BibitemOpen
  \bibfield{author}{%
  \bibinfo {author} {\bibfnamefont{Steven}\ \bibnamefont{Weinberg}},\ }%
  \bibfield{title}{%
  \enquote{\bibinfo {title} {{Infrared photons and gravitons}},}\ }%
  \bibfield{journal}{%
  \Doi{10.1103/PhysRev.140.B516}{\bibinfo {journal} {Phys. Rev.}}\ }%
  \textbf{\bibinfo {volume} {140}},\ \bibinfo {pages} {B516--B524} (\bibinfo {year} {1965})%
  \bibAnnoteFile{NoStop}{Weinberg:1965nx}%
\bibitem{Cachazo:2014fwa}%
  \BibitemOpen
  \bibfield{author}{%
  \bibinfo {author} {\bibfnamefont{Freddy}\ \bibnamefont{Cachazo}}\ and\ \bibinfo {author} {\bibfnamefont{Andrew}\ \bibnamefont{Strominger}},\ }%
  \bibfield{title}{%
  \enquote{\bibinfo {title} {{Evidence for a New Soft Graviton Theorem}},}\ }%
   (\bibinfo {month} {4}\ \bibinfo {year} {2014}),\ \Eprint{http://arxiv.org/abs/1404.4091}{arXiv:1404.4091 [hep-th]}%
  \bibAnnoteFile{NoStop}{Cachazo:2014fwa}%
\bibitem{Schwab:2014xua}%
  \BibitemOpen
  \bibfield{author}{%
  \bibinfo {author} {\bibfnamefont{Burkhard U.~W.}\ \bibnamefont{Schwab}}\ and\ \bibinfo {author} {\bibfnamefont{Anastasia}\ \bibnamefont{Volovich}},\ }%
  \bibfield{title}{%
  \enquote{\bibinfo {title} {{Subleading Soft Theorem in Arbitrary Dimensions from Scattering Equations}},}\ }%
  \bibfield{journal}{%
  \Doi{10.1103/PhysRevLett.113.101601}{\bibinfo {journal} {Phys. Rev. Lett.}}\ }%
  \textbf{\bibinfo {volume} {113}},\ \bibinfo {pages} {101601} (\bibinfo {year} {2014}),\ \Eprint{http://arxiv.org/abs/1404.7749}{arXiv:1404.7749 [hep-th]}%
  \bibAnnoteFile{NoStop}{Schwab:2014xua}%
\bibitem{Afkhami-Jeddi:2014fia}%
  \BibitemOpen
  \bibfield{author}{%
  \bibinfo {author} {\bibfnamefont{Nima}\ \bibnamefont{Afkhami-Jeddi}},\ }%
  \bibfield{title}{%
  \enquote{\bibinfo {title} {{Soft Graviton Theorem in Arbitrary Dimensions}},}\ }%
   (\bibinfo {month} {5}\ \bibinfo {year} {2014}),\ \Eprint{http://arxiv.org/abs/1405.3533}{arXiv:1405.3533 [hep-th]}%
  \bibAnnoteFile{NoStop}{Afkhami-Jeddi:2014fia}%
\bibitem{Kalousios:2014uva}%
  \BibitemOpen
  \bibfield{author}{%
  \bibinfo {author} {\bibfnamefont{Chrysostomos}\ \bibnamefont{Kalousios}}\ and\ \bibinfo {author} {\bibfnamefont{Francisco}\ \bibnamefont{Rojas}},\ }%
  \bibfield{title}{%
  \enquote{\bibinfo {title} {{Next to subleading soft-graviton theorem in arbitrary dimensions}},}\ }%
  \bibfield{journal}{%
  \Doi{10.1007/JHEP01(2015)107}{\bibinfo {journal} {JHEP}}\ }%
  \textbf{\bibinfo {volume} {01}},\ \bibinfo {pages} {107} (\bibinfo {year} {2015}),\ \Eprint{http://arxiv.org/abs/1407.5982}{arXiv:1407.5982 [hep-th]}%
  \bibAnnoteFile{NoStop}{Kalousios:2014uva}%
\bibitem{Cachazo:2014dia}%
  \BibitemOpen
  \bibfield{author}{%
  \bibinfo {author} {\bibfnamefont{Freddy}\ \bibnamefont{Cachazo}}\ and\ \bibinfo {author} {\bibfnamefont{Ellis~Ye}\ \bibnamefont{Yuan}},\ }%
  \bibfield{title}{%
  \enquote{\bibinfo {title} {{Are Soft Theorems Renormalized?}}.}\ }%
   (\bibinfo {month} {5}\ \bibinfo {year} {2014}),\ \Eprint{http://arxiv.org/abs/1405.3413}{arXiv:1405.3413 [hep-th]}%
  \bibAnnoteFile{NoStop}{Cachazo:2014dia}%
\bibitem{Bern:2014oka}%
  \BibitemOpen
  \bibfield{author}{%
  \bibinfo {author} {\bibfnamefont{Zvi}\ \bibnamefont{Bern}}, \bibinfo {author} {\bibfnamefont{Scott}\ \bibnamefont{Davies}},\ and\ \bibinfo {author} {\bibfnamefont{Josh}\ \bibnamefont{Nohle}},\ }%
  \bibfield{title}{%
  \enquote{\bibinfo {title} {{On Loop Corrections to Subleading Soft Behavior of Gluons and Gravitons}},}\ }%
  \bibfield{journal}{%
  \Doi{10.1103/PhysRevD.90.085015}{\bibinfo {journal} {Phys. Rev. D}}\ }%
  \textbf{\bibinfo {volume} {90}},\ \bibinfo {pages} {085015} (\bibinfo {year} {2014}),\ \Eprint{http://arxiv.org/abs/1405.1015}{arXiv:1405.1015 [hep-th]}%
  \bibAnnoteFile{NoStop}{Bern:2014oka}%
\bibitem{Broedel:2014fsa}%
  \BibitemOpen
  \bibfield{author}{%
  \bibinfo {author} {\bibfnamefont{Johannes}\ \bibnamefont{Broedel}}, \bibinfo {author} {\bibfnamefont{Marius}\ \bibnamefont{de~Leeuw}}, \bibinfo {author} {\bibfnamefont{Jan}\ \bibnamefont{Plefka}},\ and\ \bibinfo {author} {\bibfnamefont{Matteo}\ \bibnamefont{Rosso}},\ }%
  \bibfield{title}{%
  \enquote{\bibinfo {title} {{Constraining subleading soft gluon and graviton theorems}},}\ }%
  \bibfield{journal}{%
  \Doi{10.1103/PhysRevD.90.065024}{\bibinfo {journal} {Phys. Rev. D}}\ }%
  \textbf{\bibinfo {volume} {90}},\ \bibinfo {pages} {065024} (\bibinfo {year} {2014}),\ \Eprint{http://arxiv.org/abs/1406.6574}{arXiv:1406.6574 [hep-th]}%
  \bibAnnoteFile{NoStop}{Broedel:2014fsa}%
\bibitem{He:2014bga}%
  \BibitemOpen
  \bibfield{author}{%
  \bibinfo {author} {\bibfnamefont{Song}\ \bibnamefont{He}}, \bibinfo {author} {\bibfnamefont{Yu-tin}\ \bibnamefont{Huang}},\ and\ \bibinfo {author} {\bibfnamefont{Congkao}\ \bibnamefont{Wen}},\ }%
  \bibfield{title}{%
  \enquote{\bibinfo {title} {{Loop Corrections to Soft Theorems in Gauge Theories and Gravity}},}\ }%
  \bibfield{journal}{%
  \Doi{10.1007/JHEP12(2014)115}{\bibinfo {journal} {JHEP}}\ }%
  \textbf{\bibinfo {volume} {12}},\ \bibinfo {pages} {115} (\bibinfo {year} {2014}),\ \Eprint{http://arxiv.org/abs/1405.1410}{arXiv:1405.1410 [hep-th]}%
  \bibAnnoteFile{NoStop}{He:2014bga}%
\bibitem{Elvang:2016qvq}%
  \BibitemOpen
  \bibfield{author}{%
  \bibinfo {author} {\bibfnamefont{Henriette}\ \bibnamefont{Elvang}}, \bibinfo {author} {\bibfnamefont{Callum R.~T.}\ \bibnamefont{Jones}},\ and\ \bibinfo {author} {\bibfnamefont{Stephen~G.}\ \bibnamefont{Naculich}},\ }%
  \bibfield{title}{%
  \enquote{\bibinfo {title} {{Soft Photon and Graviton Theorems in Effective Field Theory}},}\ }%
  \bibfield{journal}{%
  \Doi{10.1103/PhysRevLett.118.231601}{\bibinfo {journal} {Phys. Rev. Lett.}}\ }%
  \textbf{\bibinfo {volume} {118}},\ \bibinfo {pages} {231601} (\bibinfo {year} {2017}),\ \Eprint{http://arxiv.org/abs/1611.07534}{arXiv:1611.07534 [hep-th]}%
  \bibAnnoteFile{NoStop}{Elvang:2016qvq}%
\bibitem{Sen:2017nim}%
  \BibitemOpen
  \bibfield{author}{%
  \bibinfo {author} {\bibfnamefont{Ashoke}\ \bibnamefont{Sen}},\ }%
  \bibfield{title}{%
  \enquote{\bibinfo {title} {{Subleading Soft Graviton Theorem for Loop Amplitudes}},}\ }%
  \bibfield{journal}{%
  \Doi{10.1007/JHEP11(2017)123}{\bibinfo {journal} {JHEP}}\ }%
  \textbf{\bibinfo {volume} {11}},\ \bibinfo {pages} {123} (\bibinfo {year} {2017}),\ \Eprint{http://arxiv.org/abs/1703.00024}{arXiv:1703.00024 [hep-th]}%
  \bibAnnoteFile{NoStop}{Sen:2017nim}%
\bibitem{Saha:2019tub}%
  \BibitemOpen
  \bibfield{author}{%
  \bibinfo {author} {\bibfnamefont{Arnab~Priya}\ \bibnamefont{Saha}}, \bibinfo {author} {\bibfnamefont{Biswajit}\ \bibnamefont{Sahoo}},\ and\ \bibinfo {author} {\bibfnamefont{Ashoke}\ \bibnamefont{Sen}},\ }%
  \bibfield{title}{%
  \enquote{\bibinfo {title} {{Proof of the classical soft graviton theorem in $D$ = 4}},}\ }%
  \bibfield{journal}{%
  \Doi{10.1007/JHEP06(2020)153}{\bibinfo {journal} {JHEP}}\ }%
  \textbf{\bibinfo {volume} {06}},\ \bibinfo {pages} {153} (\bibinfo {year} {2020}),\ \Eprint{http://arxiv.org/abs/1912.06413}{arXiv:1912.06413 [hep-th]}%
  \bibAnnoteFile{NoStop}{Saha:2019tub}%
\bibitem{Yan:2023rjh}%
  \BibitemOpen
  \bibfield{author}{%
  \bibinfo {author} {\bibfnamefont{Cynthia}\ \bibnamefont{Yan}},\ }%
  \bibfield{title}{%
  \enquote{\bibinfo {title} {{More on Torus Wormholes in 3d Gravity}},}\ }%
   (\bibinfo {month} {5}\ \bibinfo {year} {2023}),\ \Eprint{http://arxiv.org/abs/2305.10494}{arXiv:2305.10494 [hep-th]}%
  \bibAnnoteFile{NoStop}{Yan:2023rjh}%
\bibitem{Kapec:2014opa}%
  \BibitemOpen
  \bibfield{author}{%
  \bibinfo {author} {\bibfnamefont{Daniel}\ \bibnamefont{Kapec}}, \bibinfo {author} {\bibfnamefont{Vyacheslav}\ \bibnamefont{Lysov}}, \bibinfo {author} {\bibfnamefont{Sabrina}\ \bibnamefont{Pasterski}},\ and\ \bibinfo {author} {\bibfnamefont{Andrew}\ \bibnamefont{Strominger}},\ }%
  \bibfield{title}{%
  \enquote{\bibinfo {title} {{Semiclassical Virasoro symmetry of the quantum gravity $ \mathcal{S}$-matrix}},}\ }%
  \bibfield{journal}{%
  \Doi{10.1007/JHEP08(2014)058}{\bibinfo {journal} {JHEP}}\ }%
  \textbf{\bibinfo {volume} {08}},\ \bibinfo {pages} {058} (\bibinfo {year} {2014}),\ \Eprint{http://arxiv.org/abs/1406.3312}{arXiv:1406.3312 [hep-th]}%
  \bibAnnoteFile{NoStop}{Kapec:2014opa}%
\bibitem{Kapec:2015vwa}%
  \BibitemOpen
  \bibfield{author}{%
  \bibinfo {author} {\bibfnamefont{Daniel}\ \bibnamefont{Kapec}}, \bibinfo {author} {\bibfnamefont{Vyacheslav}\ \bibnamefont{Lysov}}, \bibinfo {author} {\bibfnamefont{Sabrina}\ \bibnamefont{Pasterski}},\ and\ \bibinfo {author} {\bibfnamefont{Andrew}\ \bibnamefont{Strominger}},\ }%
  \bibfield{title}{%
  \enquote{\bibinfo {title} {{Higher-dimensional supertranslations and Weinberg\textquoteright{}s soft graviton theorem}},}\ }%
  \bibfield{journal}{%
  \Doi{10.4310/AMSA.2017.v2.n1.a2}{\bibinfo {journal} {Ann. Math. Sci. Appl.}}\ }%
  \textbf{\bibinfo {volume} {02}},\ \bibinfo {pages} {69--94} (\bibinfo {year} {2017}),\ \Eprint{http://arxiv.org/abs/1502.07644}{arXiv:1502.07644 [gr-qc]}%
  \bibAnnoteFile{NoStop}{Kapec:2015vwa}%
\bibitem{Strominger:2017zoo}%
  \BibitemOpen
  \bibfield{author}{%
  \bibinfo {author} {\bibfnamefont{Andrew}\ \bibnamefont{Strominger}},\ }%
  \emph{\bibinfo {title} {{Lectures on the Infrared Structure of Gravity and Gauge Theory}}}\ (\bibinfo {year} {2017})\ ISBN \bibinfo {isbn} {978-0-691-17973-5},\ \Eprint{http://arxiv.org/abs/1703.05448}{arXiv:1703.05448 [hep-th]}%
  \bibAnnoteFile{NoStop}{Strominger:2017zoo}%
\bibitem{Miller:2022fvc}%
  \BibitemOpen
  \bibfield{author}{%
  \bibinfo {author} {\bibfnamefont{Noah}\ \bibnamefont{Miller}}, \bibinfo {author} {\bibfnamefont{Andrew}\ \bibnamefont{Strominger}}, \bibinfo {author} {\bibfnamefont{Adam}\ \bibnamefont{Tropper}},\ and\ \bibinfo {author} {\bibfnamefont{Tianli}\ \bibnamefont{Wang}},\ }%
  \bibfield{title}{%
  \enquote{\bibinfo {title} {{Soft gravitons in the BFSS matrix model}},}\ }%
  \bibfield{journal}{%
  \Doi{10.1007/JHEP11(2023)174}{\bibinfo {journal} {JHEP}}\ }%
  \textbf{\bibinfo {volume} {11}},\ \bibinfo {pages} {174} (\bibinfo {year} {2023}),\ \Eprint{http://arxiv.org/abs/2208.14547}{arXiv:2208.14547 [hep-th]}%
  \bibAnnoteFile{NoStop}{Miller:2022fvc}%
\bibitem{Tropper:2023fjr}%
  \BibitemOpen
  \bibfield{author}{%
  \bibinfo {author} {\bibfnamefont{Adam}\ \bibnamefont{Tropper}}\ and\ \bibinfo {author} {\bibfnamefont{Tianli}\ \bibnamefont{Wang}},\ }%
  \bibfield{title}{%
  \enquote{\bibinfo {title} {{Lorentz symmetry and IR structure of the BFSS matrix model}},}\ }%
  \bibfield{journal}{%
  \Doi{10.1007/JHEP07(2023)150}{\bibinfo {journal} {JHEP}}\ }%
  \textbf{\bibinfo {volume} {07}},\ \bibinfo {pages} {150} (\bibinfo {year} {2023}),\ \Eprint{http://arxiv.org/abs/2303.14200}{arXiv:2303.14200 [hep-th]}%
  \bibAnnoteFile{NoStop}{Tropper:2023fjr}%
\bibitem{Herderschee:2023pza}%
  \BibitemOpen
  \bibfield{author}{%
  \bibinfo {author} {\bibfnamefont{Aidan}\ \bibnamefont{Herderschee}}\ and\ \bibinfo {author} {\bibfnamefont{Juan}\ \bibnamefont{Maldacena}},\ }%
  \bibfield{title}{%
  \enquote{\bibinfo {title} {{Three Point Amplitudes in Matrix Theory}},}\ }%
   (\bibinfo {month} {12}\ \bibinfo {year} {2023}),\ \Eprint{http://arxiv.org/abs/2312.12592}{arXiv:2312.12592 [hep-th]}%
  \bibAnnoteFile{NoStop}{Herderschee:2023pza}%
\bibitem{Elvang:2013cua}%
  \BibitemOpen
  \bibfield{author}{%
  \bibinfo {author} {\bibfnamefont{Henriette}\ \bibnamefont{Elvang}}\ and\ \bibinfo {author} {\bibfnamefont{Yu-tin}\ \bibnamefont{Huang}},\ }%
  \bibfield{title}{%
  \enquote{\bibinfo {title} {{Scattering Amplitudes}},}\ }%
   (\bibinfo {month} {8}\ \bibinfo {year} {2013}),\ \Eprint{http://arxiv.org/abs/1308.1697}{arXiv:1308.1697 [hep-th]}%
  \bibAnnoteFile{NoStop}{Elvang:2013cua}%
\bibitem{Britto:2004ap}%
  \BibitemOpen
  \bibfield{author}{%
  \bibinfo {author} {\bibfnamefont{Ruth}\ \bibnamefont{Britto}}, \bibinfo {author} {\bibfnamefont{Freddy}\ \bibnamefont{Cachazo}},\ and\ \bibinfo {author} {\bibfnamefont{Bo}~\bibnamefont{Feng}},\ }%
  \bibfield{title}{%
  \enquote{\bibinfo {title} {{New recursion relations for tree amplitudes of gluons}},}\ }%
  \bibfield{journal}{%
  \Doi{10.1016/j.nuclphysb.2005.02.030}{\bibinfo {journal} {Nucl. Phys. B}}\ }%
  \textbf{\bibinfo {volume} {715}},\ \bibinfo {pages} {499--522} (\bibinfo {year} {2005}),\ \Eprint{http://arxiv.org/abs/hep-th/0412308}{arXiv:hep-th/0412308}%
  \bibAnnoteFile{NoStop}{Britto:2004ap}%
\bibitem{Britto:2005fq}%
  \BibitemOpen
  \bibfield{author}{%
  \bibinfo {author} {\bibfnamefont{Ruth}\ \bibnamefont{Britto}}, \bibinfo {author} {\bibfnamefont{Freddy}\ \bibnamefont{Cachazo}}, \bibinfo {author} {\bibfnamefont{Bo}~\bibnamefont{Feng}},\ and\ \bibinfo {author} {\bibfnamefont{Edward}\ \bibnamefont{Witten}},\ }%
  \bibfield{title}{%
  \enquote{\bibinfo {title} {{Direct proof of tree-level recursion relation in Yang-Mills theory}},}\ }%
  \bibfield{journal}{%
  \Doi{10.1103/PhysRevLett.94.181602}{\bibinfo {journal} {Phys. Rev. Lett.}}\ }%
  \textbf{\bibinfo {volume} {94}},\ \bibinfo {pages} {181602} (\bibinfo {year} {2005}),\ \Eprint{http://arxiv.org/abs/hep-th/0501052}{arXiv:hep-th/0501052}%
  \bibAnnoteFile{NoStop}{Britto:2005fq}%
\bibitem{Arkani-Hamed:2008owk}%
  \BibitemOpen
  \bibfield{author}{%
  \bibinfo {author} {\bibfnamefont{Nima}\ \bibnamefont{Arkani-Hamed}}, \bibinfo {author} {\bibfnamefont{Freddy}\ \bibnamefont{Cachazo}},\ and\ \bibinfo {author} {\bibfnamefont{Jared}\ \bibnamefont{Kaplan}},\ }%
  \bibfield{title}{%
  \enquote{\bibinfo {title} {{What is the Simplest Quantum Field Theory?}}.}\ }%
  \bibfield{journal}{%
  \Doi{10.1007/JHEP09(2010)016}{\bibinfo {journal} {JHEP}}\ }%
  \textbf{\bibinfo {volume} {09}},\ \bibinfo {pages} {016} (\bibinfo {year} {2010}),\ \Eprint{http://arxiv.org/abs/0808.1446}{arXiv:0808.1446 [hep-th]}%
  \bibAnnoteFile{NoStop}{Arkani-Hamed:2008owk}%
\bibitem{Risager:2005vk}%
  \BibitemOpen
  \bibfield{author}{%
  \bibinfo {author} {\bibfnamefont{Kasper}\ \bibnamefont{Risager}},\ }%
  \bibfield{title}{%
  \enquote{\bibinfo {title} {{A Direct proof of the CSW rules}},}\ }%
  \bibfield{journal}{%
  \Doi{10.1088/1126-6708/2005/12/003}{\bibinfo {journal} {JHEP}}\ }%
  \textbf{\bibinfo {volume} {12}},\ \bibinfo {pages} {003} (\bibinfo {year} {2005}),\ \Eprint{http://arxiv.org/abs/hep-th/0508206}{arXiv:hep-th/0508206}%
  \bibAnnoteFile{NoStop}{Risager:2005vk}%
\bibitem{Bjerrum-Bohr:2005xoa}%
  \BibitemOpen
  \bibfield{author}{%
  \bibinfo {author} {\bibfnamefont{N.~E.~J.}\ \bibnamefont{Bjerrum-Bohr}}, \bibinfo {author} {\bibfnamefont{David~C.}\ \bibnamefont{Dunbar}}, \bibinfo {author} {\bibfnamefont{Harald}\ \bibnamefont{Ita}}, \bibinfo {author} {\bibfnamefont{Warren~B.}\ \bibnamefont{Perkins}},\ and\ \bibinfo {author} {\bibfnamefont{Kasper}\ \bibnamefont{Risager}},\ }%
  \bibfield{title}{%
  \enquote{\bibinfo {title} {{MHV-vertices for gravity amplitudes}},}\ }%
  \bibfield{journal}{%
  \Doi{10.1088/1126-6708/2006/01/009}{\bibinfo {journal} {JHEP}}\ }%
  \textbf{\bibinfo {volume} {01}},\ \bibinfo {pages} {009} (\bibinfo {year} {2006}),\ \Eprint{http://arxiv.org/abs/hep-th/0509016}{arXiv:hep-th/0509016}%
  \bibAnnoteFile{NoStop}{Bjerrum-Bohr:2005xoa}%
\bibitem{Bianchi:2008pu}%
  \BibitemOpen
  \bibfield{author}{%
  \bibinfo {author} {\bibfnamefont{Massimo}\ \bibnamefont{Bianchi}}, \bibinfo {author} {\bibfnamefont{Henriette}\ \bibnamefont{Elvang}},\ and\ \bibinfo {author} {\bibfnamefont{Daniel~Z.}\ \bibnamefont{Freedman}},\ }%
  \bibfield{title}{%
  \enquote{\bibinfo {title} {{Generating Tree Amplitudes in N=4 SYM and N = 8 SG}},}\ }%
  \bibfield{journal}{%
  \Doi{10.1088/1126-6708/2008/09/063}{\bibinfo {journal} {JHEP}}\ }%
  \textbf{\bibinfo {volume} {09}},\ \bibinfo {pages} {063} (\bibinfo {year} {2008}),\ \Eprint{http://arxiv.org/abs/0805.0757}{arXiv:0805.0757 [hep-th]}%
  \bibAnnoteFile{NoStop}{Bianchi:2008pu}%
\bibitem{Cohen:2010mi}%
  \BibitemOpen
  \bibfield{author}{%
  \bibinfo {author} {\bibfnamefont{Timothy}\ \bibnamefont{Cohen}}, \bibinfo {author} {\bibfnamefont{Henriette}\ \bibnamefont{Elvang}},\ and\ \bibinfo {author} {\bibfnamefont{Michael}\ \bibnamefont{Kiermaier}},\ }%
  \bibfield{title}{%
  \enquote{\bibinfo {title} {{On-shell constructibility of tree amplitudes in general field theories}},}\ }%
  \bibfield{journal}{%
  \Doi{10.1007/JHEP04(2011)053}{\bibinfo {journal} {JHEP}}\ }%
  \textbf{\bibinfo {volume} {04}},\ \bibinfo {pages} {053} (\bibinfo {year} {2011}),\ \Eprint{http://arxiv.org/abs/1010.0257}{arXiv:1010.0257 [hep-th]}%
  \bibAnnoteFile{NoStop}{Cohen:2010mi}%
\bibitem{Yi:1997eg}%
  \BibitemOpen
  \bibfield{author}{%
  \bibinfo {author} {\bibfnamefont{Piljin}\ \bibnamefont{Yi}},\ }%
  \bibfield{title}{%
  \enquote{\bibinfo {title} {{Witten index and threshold bound states of D-branes}},}\ }%
  \bibfield{journal}{%
  \Doi{10.1016/S0550-3213(97)00486-0}{\bibinfo {journal} {Nucl. Phys. B}}\ }%
  \textbf{\bibinfo {volume} {505}},\ \bibinfo {pages} {307--318} (\bibinfo {year} {1997}),\ \Eprint{http://arxiv.org/abs/hep-th/9704098}{arXiv:hep-th/9704098}%
  \bibAnnoteFile{NoStop}{Yi:1997eg}%
\bibitem{Sethi:1997pa}%
  \BibitemOpen
  \bibfield{author}{%
  \bibinfo {author} {\bibfnamefont{Savdeep}\ \bibnamefont{Sethi}}\ and\ \bibinfo {author} {\bibfnamefont{Mark}\ \bibnamefont{Stern}},\ }%
  \bibfield{title}{%
  \enquote{\bibinfo {title} {{D-brane bound states redux}},}\ }%
  \bibfield{journal}{%
  \Doi{10.1007/s002200050374}{\bibinfo {journal} {Commun. Math. Phys.}}\ }%
  \textbf{\bibinfo {volume} {194}},\ \bibinfo {pages} {675--705} (\bibinfo {year} {1998}),\ \Eprint{http://arxiv.org/abs/hep-th/9705046}{arXiv:hep-th/9705046}%
  \bibAnnoteFile{NoStop}{Sethi:1997pa}%
\bibitem{Moore:1998et}%
  \BibitemOpen
  \bibfield{author}{%
  \bibinfo {author} {\bibfnamefont{Gregory~W.}\ \bibnamefont{Moore}}, \bibinfo {author} {\bibfnamefont{Nikita}\ \bibnamefont{Nekrasov}},\ and\ \bibinfo {author} {\bibfnamefont{Samson}\ \bibnamefont{Shatashvili}},\ }%
  \bibfield{title}{%
  \enquote{\bibinfo {title} {{D particle bound states and generalized instantons}},}\ }%
  \bibfield{journal}{%
  \Doi{10.1007/s002200050016}{\bibinfo {journal} {Commun. Math. Phys.}}\ }%
  \textbf{\bibinfo {volume} {209}},\ \bibinfo {pages} {77--95} (\bibinfo {year} {2000}),\ \Eprint{http://arxiv.org/abs/hep-th/9803265}{arXiv:hep-th/9803265}%
  \bibAnnoteFile{NoStop}{Moore:1998et}%
\bibitem{Konechny:1998vc}%
  \BibitemOpen
  \bibfield{author}{%
  \bibinfo {author} {\bibfnamefont{Anatoly}\ \bibnamefont{Konechny}},\ }%
  \bibfield{title}{%
  \enquote{\bibinfo {title} {{On asymptotic Hamiltonian for SU(N) matrix theory}},}\ }%
  \bibfield{journal}{%
  \Doi{10.1088/1126-6708/1998/10/018}{\bibinfo {journal} {JHEP}}\ }%
  \textbf{\bibinfo {volume} {10}},\ \bibinfo {pages} {018} (\bibinfo {year} {1998}),\ \Eprint{http://arxiv.org/abs/hep-th/9805046}{arXiv:hep-th/9805046}%
  \bibAnnoteFile{NoStop}{Konechny:1998vc}%
\bibitem{Porrati:1997ej}%
  \BibitemOpen
  \bibfield{author}{%
  \bibinfo {author} {\bibfnamefont{Massimo}\ \bibnamefont{Porrati}}\ and\ \bibinfo {author} {\bibfnamefont{Alexander}\ \bibnamefont{Rozenberg}},\ }%
  \bibfield{title}{%
  \enquote{\bibinfo {title} {{Bound states at threshold in supersymmetric quantum mechanics}},}\ }%
  \bibfield{journal}{%
  \Doi{10.1016/S0550-3213(97)00804-3}{\bibinfo {journal} {Nucl. Phys. B}}\ }%
  \textbf{\bibinfo {volume} {515}},\ \bibinfo {pages} {184--202} (\bibinfo {year} {1998}),\ \Eprint{http://arxiv.org/abs/hep-th/9708119}{arXiv:hep-th/9708119}%
  \bibAnnoteFile{NoStop}{Porrati:1997ej}%
\bibitem{Sethi:2000zf}%
  \BibitemOpen
  \bibfield{author}{%
  \bibinfo {author} {\bibfnamefont{Savdeep}\ \bibnamefont{Sethi}}\ and\ \bibinfo {author} {\bibfnamefont{Mark}\ \bibnamefont{Stern}},\ }%
  \bibfield{title}{%
  \enquote{\bibinfo {title} {{Invariance theorems for supersymmetric Yang-Mills theories}},}\ }%
  \bibfield{journal}{%
  \Doi{10.4310/ATMP.2000.v4.n2.a8}{\bibinfo {journal} {Adv. Theor. Math. Phys.}}\ }%
  \textbf{\bibinfo {volume} {4}},\ \bibinfo {pages} {487--501} (\bibinfo {year} {2000}),\ \Eprint{http://arxiv.org/abs/hep-th/0001189}{arXiv:hep-th/0001189}%
  \bibAnnoteFile{NoStop}{Sethi:2000zf}%
\bibitem{Lin:2014wka}%
  \BibitemOpen
  \bibfield{author}{%
  \bibinfo {author} {\bibfnamefont{Ying-Hsuan}\ \bibnamefont{Lin}}\ and\ \bibinfo {author} {\bibfnamefont{Xi}~\bibnamefont{Yin}},\ }%
  \bibfield{title}{%
  \enquote{\bibinfo {title} {{On the Ground State Wave Function of Matrix Theory}},}\ }%
  \bibfield{journal}{%
  \Doi{10.1007/JHEP11(2015)027}{\bibinfo {journal} {JHEP}}\ }%
  \textbf{\bibinfo {volume} {11}},\ \bibinfo {pages} {027} (\bibinfo {year} {2015}),\ \Eprint{http://arxiv.org/abs/1402.0055}{arXiv:1402.0055 [hep-th]}%
  \bibAnnoteFile{NoStop}{Lin:2014wka}%
\bibitem{deWit:1988xki}%
  \BibitemOpen
  \bibfield{author}{%
  \bibinfo {author} {\bibfnamefont{B.}~\bibnamefont{de~Wit}}, \bibinfo {author} {\bibfnamefont{M.}~\bibnamefont{Luscher}},\ and\ \bibinfo {author} {\bibfnamefont{H.}~\bibnamefont{Nicolai}},\ }%
  \bibfield{title}{%
  \enquote{\bibinfo {title} {{The Supermembrane Is Unstable}},}\ }%
  \bibfield{journal}{%
  \Doi{10.1016/0550-3213(89)90214-9}{\bibinfo {journal} {Nucl. Phys. B}}\ }%
  \textbf{\bibinfo {volume} {320}},\ \bibinfo {pages} {135--159} (\bibinfo {year} {1989})%
  \bibAnnoteFile{NoStop}{deWit:1988xki}%
\bibitem{Bern:2014vva}%
  \BibitemOpen
  \bibfield{author}{%
  \bibinfo {author} {\bibfnamefont{Zvi}\ \bibnamefont{Bern}}, \bibinfo {author} {\bibfnamefont{Scott}\ \bibnamefont{Davies}}, \bibinfo {author} {\bibfnamefont{Paolo}\ \bibnamefont{Di~Vecchia}},\ and\ \bibinfo {author} {\bibfnamefont{Josh}\ \bibnamefont{Nohle}},\ }%
  \bibfield{title}{%
  \enquote{\bibinfo {title} {{Low-Energy Behavior of Gluons and Gravitons from Gauge Invariance}},}\ }%
  \bibfield{journal}{%
  \Doi{10.1103/PhysRevD.90.084035}{\bibinfo {journal} {Phys. Rev. D}}\ }%
  \textbf{\bibinfo {volume} {90}},\ \bibinfo {pages} {084035} (\bibinfo {year} {2014}),\ \Eprint{http://arxiv.org/abs/1406.6987}{arXiv:1406.6987 [hep-th]}%
  \bibAnnoteFile{NoStop}{Bern:2014vva}%
\bibitem{Krishna:2023fxg}%
  \BibitemOpen
  \bibfield{author}{%
  \bibinfo {author} {\bibfnamefont{Hare}\ \bibnamefont{Krishna}}\ and\ \bibinfo {author} {\bibfnamefont{Biswajit}\ \bibnamefont{Sahoo}},\ }%
  \bibfield{title}{%
  \enquote{\bibinfo {title} {{Universality of loop corrected soft theorems in 4d}},}\ }%
  \bibfield{journal}{%
  \Doi{10.1007/JHEP11(2023)233}{\bibinfo {journal} {JHEP}}\ }%
  \textbf{\bibinfo {volume} {11}},\ \bibinfo {pages} {233} (\bibinfo {year} {2023}),\ \Eprint{http://arxiv.org/abs/2308.16807}{arXiv:2308.16807 [hep-th]}%
  \bibAnnoteFile{NoStop}{Krishna:2023fxg}%
\bibitem{Gary:2009ae}%
  \BibitemOpen
  \bibfield{author}{%
  \bibinfo {author} {\bibfnamefont{Mirah}\ \bibnamefont{Gary}}, \bibinfo {author} {\bibfnamefont{Steven~B.}\ \bibnamefont{Giddings}},\ and\ \bibinfo {author} {\bibfnamefont{Joao}\ \bibnamefont{Penedones}},\ }%
  \bibfield{title}{%
  \enquote{\bibinfo {title} {{Local bulk S-matrix elements and CFT singularities}},}\ }%
  \bibfield{journal}{%
  \Doi{10.1103/PhysRevD.80.085005}{\bibinfo {journal} {Phys. Rev. D}}\ }%
  \textbf{\bibinfo {volume} {80}},\ \bibinfo {pages} {085005} (\bibinfo {year} {2009}),\ \Eprint{http://arxiv.org/abs/0903.4437}{arXiv:0903.4437 [hep-th]}%
  \bibAnnoteFile{NoStop}{Gary:2009ae}%
\bibitem{Maldacena:2015iua}%
  \BibitemOpen
  \bibfield{author}{%
  \bibinfo {author} {\bibfnamefont{Juan}\ \bibnamefont{Maldacena}}, \bibinfo {author} {\bibfnamefont{David}\ \bibnamefont{Simmons-Duffin}},\ and\ \bibinfo {author} {\bibfnamefont{Alexander}\ \bibnamefont{Zhiboedov}},\ }%
  \bibfield{title}{%
  \enquote{\bibinfo {title} {{Looking for a bulk point}},}\ }%
  \bibfield{journal}{%
  \Doi{10.1007/JHEP01(2017)013}{\bibinfo {journal} {JHEP}}\ }%
  \textbf{\bibinfo {volume} {01}},\ \bibinfo {pages} {013} (\bibinfo {year} {2017}),\ \Eprint{http://arxiv.org/abs/1509.03612}{arXiv:1509.03612 [hep-th]}%
  \bibAnnoteFile{NoStop}{Maldacena:2015iua}%
\bibitem{Itzhaki:1998dd}%
  \BibitemOpen
  \bibfield{author}{%
  \bibinfo {author} {\bibfnamefont{Nissan}\ \bibnamefont{Itzhaki}}, \bibinfo {author} {\bibfnamefont{Juan~Martin}\ \bibnamefont{Maldacena}}, \bibinfo {author} {\bibfnamefont{Jacob}\ \bibnamefont{Sonnenschein}},\ and\ \bibinfo {author} {\bibfnamefont{Shimon}\ \bibnamefont{Yankielowicz}},\ }%
  \bibfield{title}{%
  \enquote{\bibinfo {title} {{Supergravity and the large N limit of theories with sixteen supercharges}},}\ }%
  \bibfield{journal}{%
  \Doi{10.1103/PhysRevD.58.046004}{\bibinfo {journal} {Phys. Rev. D}}\ }%
  \textbf{\bibinfo {volume} {58}},\ \bibinfo {pages} {046004} (\bibinfo {year} {1998}),\ \Eprint{http://arxiv.org/abs/hep-th/9802042}{arXiv:hep-th/9802042}%
  \bibAnnoteFile{NoStop}{Itzhaki:1998dd}%
\bibitem{Bergner:2021goh}%
  \BibitemOpen
  \bibfield{author}{%
  \bibinfo {author} {\bibfnamefont{Georg}\ \bibnamefont{Bergner}}, \bibinfo {author} {\bibfnamefont{Norbert}\ \bibnamefont{Bodendorfer}}, \bibinfo {author} {\bibfnamefont{Masanori}\ \bibnamefont{Hanada}}, \bibinfo {author} {\bibfnamefont{Stratos}\ \bibnamefont{Pateloudis}}, \bibinfo {author} {\bibfnamefont{Enrico}\ \bibnamefont{Rinaldi}}, \bibinfo {author} {\bibfnamefont{Andreas}\ \bibnamefont{Sch\"afer}}, \bibinfo {author} {\bibfnamefont{Pavlos}\ \bibnamefont{Vranas}},\ and\ \bibinfo {author} {\bibfnamefont{Hiromasa}\ \bibnamefont{Watanabe}} (\bibinfo {collaboration} {MCSMC}),\ }%
  \bibfield{title}{%
  \enquote{\bibinfo {title} {{Confinement/deconfinement transition in the D0-brane matrix model -- A signature of M-theory?}}.}\ }%
  \bibfield{journal}{%
  \Doi{10.1007/JHEP05(2022)096}{\bibinfo {journal} {JHEP}}\ }%
  \textbf{\bibinfo {volume} {05}},\ \bibinfo {pages} {096} (\bibinfo {year} {2022}),\ \Eprint{http://arxiv.org/abs/2110.01312}{arXiv:2110.01312 [hep-th]}%
  \bibAnnoteFile{NoStop}{Bergner:2021goh}%
\bibitem{Cachazo:2005ca}%
  \BibitemOpen
  \bibfield{author}{%
  \bibinfo {author} {\bibfnamefont{Freddy}\ \bibnamefont{Cachazo}}\ and\ \bibinfo {author} {\bibfnamefont{Peter}\ \bibnamefont{Svrcek}},\ }%
  \bibfield{title}{%
  \enquote{\bibinfo {title} {{Tree level recursion relations in general relativity}},}\ }%
   (\bibinfo {month} {2}\ \bibinfo {year} {2005}),\ \Eprint{http://arxiv.org/abs/hep-th/0502160}{arXiv:hep-th/0502160}%
  \bibAnnoteFile{NoStop}{Cachazo:2005ca}%
\bibitem{Bedford:2005yy}%
  \BibitemOpen
  \bibfield{author}{%
  \bibinfo {author} {\bibfnamefont{James}\ \bibnamefont{Bedford}}, \bibinfo {author} {\bibfnamefont{Andreas}\ \bibnamefont{Brandhuber}}, \bibinfo {author} {\bibfnamefont{Bill~J.}\ \bibnamefont{Spence}},\ and\ \bibinfo {author} {\bibfnamefont{Gabriele}\ \bibnamefont{Travaglini}},\ }%
  \bibfield{title}{%
  \enquote{\bibinfo {title} {{A Recursion relation for gravity amplitudes}},}\ }%
  \bibfield{journal}{%
  \Doi{10.1016/j.nuclphysb.2005.016}{\bibinfo {journal} {Nucl. Phys. B}}\ }%
  \textbf{\bibinfo {volume} {721}},\ \bibinfo {pages} {98--110} (\bibinfo {year} {2005}),\ \Eprint{http://arxiv.org/abs/hep-th/0502146}{arXiv:hep-th/0502146}%
  \bibAnnoteFile{NoStop}{Bedford:2005yy}%
\bibitem{Hodges:2011wm}%
  \BibitemOpen
  \bibfield{author}{%
  \bibinfo {author} {\bibfnamefont{Andrew}\ \bibnamefont{Hodges}},\ }%
  \bibfield{title}{%
  \enquote{\bibinfo {title} {{New expressions for gravitational scattering amplitudes}},}\ }%
  \bibfield{journal}{%
  \Doi{10.1007/JHEP07(2013)075}{\bibinfo {journal} {JHEP}}\ }%
  \textbf{\bibinfo {volume} {07}},\ \bibinfo {pages} {075} (\bibinfo {year} {2013}),\ \Eprint{http://arxiv.org/abs/1108.2227}{arXiv:1108.2227 [hep-th]}%
  \bibAnnoteFile{NoStop}{Hodges:2011wm}%
\end{thebibliography}%
\end{document}